# Feasibility of GNSS-free Localization: A TDoA-based Approach Using LoRaWAN

Thesis submitted in partial fulfillment
of the requirements for the degree of

*Master of Science*

*in*

***Electronics and Communication Engineering** by Research*

by

Ruthwik Muppala
20161061
ruthwik.muppala@research.iiit.ac.in

International Institute of Information Technology
Hyderabad - 500 032, INDIA

July 2021

International Institute of Information Technology

Hyderabad, India

# CERTIFICATE

It is certified that the work contained in this thesis, titled “Feasibility of GNSS-free Localization: A TDoA-based Approach Using LoRaWAN” by Ruthwik Muppala, has been carried out under my supervision and is not submitted elsewhere for a degree.

_____________ Date

_________________________________ Advisor: Dr. Aftab M Hussain

# Acknowledgments

I would like to express my deepest gratitude to my advisor and mentor, Dr. Aftab M. Hussain for his excellent guidance and for inspiring me to continue my path in academia. This work would not have been possible without his constant motivation. I also thank my lab mates at PATRIoT Lab, CVEST for their helpful insights, fun, and support. I am also grateful to my close friends who made this journey exciting and for always having my back. I deeply thank my mother for the constant emotional support and love. A special thanks to my brother for his vast and invaluable advice on everything.

# Abstract

The advent of the Internet of Things (IoT) has resulted in the rise of low power wide area networks such as LoRaWAN, Sigfox, and NB-IoT. Among these, LoRaWAN has garnered tremendous attention owing to the low power consumption of end nodes, long range, high resistance to multipath, low cost, and use of license-free sub-GHz bands. Consequently, LoRaWAN is gradually replacing Wi-Fi and Bluetooth in sundry IoT applications including utility metering, smart cities, and localization. Localization, in particular, has already witnessed a surge of alternatives to Global Navigation Satellite System (GNSS), based on Wi-Fi, Bluetooth, Ultra Wide Band, 5G, etc. in indoor and low power domains due to the poor indoor coverage and high power consumption of GNSS. With the need for localization only shooting up with dense IoT deployments, LoRaWAN is seen as a promising solution in this context. Indeed, many attempts employing various techniques such as Time of Arrival (ToA), Time Difference of Arrival (TDoA), and Received Signal Strength Index (RSSI) have been made to achieve localization using LoRaWAN. However, a significant drawback in this scenario is the lack of extensive data on path loss and signal propagation modeling, particularly in Indian cityscapes. Another demerit is the use of GNSS at some stage primarily for time synchronization of gateways. In this work, we attempt to nullify these two disadvantages of LoRaWAN based localization. The first part of this work presents experimental data of LoRaWAN transmissions inside a typical city building to study signal propagation and path loss. The latter part proposes a standalone GNSS-free localization approach using LoRaWAN that is achieved by applying a collaborative, TDoA-based methodology. An additional stationary node is introduced into the network to allow the synchronization of gateways without GNSS. Finally, the distribution of localization error in a triangle of gateways and the effect of timing resolution, time-on-air, and duty cycle constraints on it are investigated.

# Table of Contents

# List of Abbreviations

| | |
|---|---|
| **AoA** | Angle of Arrival |
| **BLE** | Bluetooth Low Energy |
| **BW** | Bandwidth |
| **CSS** | Chirp Spread Spectrum |
| **CR** | Coding Rate |
| **CRC** | Cyclic Redundancy Check |
| **DTDoA** | Differential Time Difference of Arrival |
| **GNSS** | Global Navigation Satellite Systems |
| **GSM** | Global System for Mobile Communication |
| **IoT** | Internet of Things |
| **IIoT** | Industrial Internet of Things |
| **LPWAN** | Low Power Wide Area Network |
| **LoRa** | Long Range |
| **LoRaWAN** | Long Range Wide Area Network |
| **LoS** | Line of Sight |
| **LSF** | Large Scale Fading |
| **MAC** | Medium Access Control |
| **NB-IoT** | Narrow Band Internet of Things |
| **OFDM** | Orthogonal Frequency Division Multiplexing |
| **OTA** | Over The Air |
| **PHY** | Physical Layer |
| **RSSI** | Received Signal Strength Index |
| **RTC** | Real Time Clock |
| **SF** | Spreading Factor |
| **SNR** | Signal to Noise Ratio |
| **ToA** | Time of Arrival |
| **TCP/IP** | Transmission Control Protocol/Internet Protocol |
| **TDoA** | Time Difference of Arrival |
| **UWB** | Ultra Wide Band |
| **WSN** | Wireless Sensor Network |

# List of Symbols

| | |
|---|---|
| $x_0$ | x-coordinate of the location of the synchronization node |
| $y_0$ | y-coordinate of the location of the synchronization node |
| $a_j$ | x-coordinate of the location of the $j^{th}$ gateway |
| $b_j$ | y-coordinate of the location of the $j^{th}$ gateway |
| $x$ | x-coordinate of the location of the target node |
| $y$ | y-coordinate of the location of the target node |
| $t_j$ | ToA value captured at the $j^{th}$ gateway |
| $c$ | Speed of light |
| $d_j$ | Distance between the $j^{th}$ gateway and the target node |
| $t_0$ | Time of emission of the target node |
| $i$ | Complex number $\sqrt{-1}$ |
| $\boldsymbol{p}$ | Position vector of the target node |
| $\boldsymbol{w_j}$ | Position vector of the $j^{th}$ gateway |
| $\alpha$ | Period of transmission of the synchronization node |
| $t_{d_j}$ | ToA value of synchronization node transmission at the $j^{th}$ gateway |
| $e_{t_j}$ | Error in $t_j$ |
| $t_{d_j}$ | Error in $t_{d_j}$ |
| $\omega_{1j}$ | Gaussian model of oscillator drift in counter ($j^{th}$ gateway) |
| $\sigma_{1j}$ | Standard deviation of the distribution $\omega_{1j}$ |
| $\omega_{2j}$ | Gaussian model of oscillator drift in processor clock ($j^{th}$ gateway) |
| $\sigma_{1j}$ | Standard deviation of the distribution $\omega_{2j}$ |
| $T/f$ | Time-period / Frequency of synchronous counter clock |
| $T_g/f_g$ | Time-period / Frequency of processor clock |
| $N_j$ | Multiple of T stored in the counter corresponding to the $t_j$ value |
| $n$ | Bit resolution of the synchronous counter |
| $\mathcal{U}_{1j}$ | Discrete uniform distribution to model processor clock slippages |
| $k_j$ | Maximum number of clock slippages |
| $\mathcal{U}_{2j}$ | Uniform distribution model of rounding off error in the $n$-bit counter |
| $\Delta x_0$ | Error in x-coordinate of the synchronization node |
| $\Delta y_0$ | Error in y-coordinate of the synchronization node |
| $\delta$ | Duty Cycle |
| $\delta_{max}$ | Maximum duty cycle allowed |
| $e_{max}$ | Maximum acceptable mean error in localization |
| $e_{t_{ideal}}$ | Error in $t_j$ in the ideal case |
| $e_{t_{ideal,max}}$ | Maximum error in $t_j$ in the ideal case |
| $\tau$ | Time on air |
| $T_{sym}$ | Symbol duration |
| $T_{preamble}$ | Preamble duration |
| $T_{payload}$ | Payload duration |
| $n_{preamble}$ | Number of symbols in the preamble |
| $n_{payload+header}$ | Total number of symbols in the payload and header |

# List of Tables

# List of Figures

*Chapter 1*

# Introduction

While IoT has been one of the central elements in diverse research fields such as smart cities [1]–[3], industrial applications (IIoT) [4]–[6], health care [7]–[10], transistor technologies [11]–[13], and flexible electronics [14]–[20] over the last decade, recent years have also witnessed its meteoric rise in consumer applications such as home automation, transportation, energy management, etc. However, many of these devices still use existing protocols such as Wi-Fi [21], Cellular [22], and Bluetooth [23]. The bandwidth and sophistication of these protocols are not required for most IoT applications. Instead, low power consumption and longer ranges take precedence. This led to the development of protocols tailor-made for IoT applications, known as LPWANs. Indeed, the chief advantages of these protocols are low power consumption, long range, and low cost. Among these, LoRaWAN has drawn plenty of attention in recent years owing to the following advantages:

i. License-free nature. It operates in unlicensed sub-GHz spectrums.
ii. The ability to choose power consumption to be low or high in line with the specific use cases [24].
iii. Potentially better battery lifetime of end devices and lower cost, especially when compared to NB-IoT [25], [26].
iv. The large coverage provided by the high link budget of the LoRa modulation protocol. Thus, fewer gateways are required to provide LoRaWAN coverage over a relatively large area.
v. High resilience to multipath fading over long distances [27], [28].

Consequently, LoRaWAN is gradually replacing existing protocols in a slew of IoT applications including smart utility metering [29], [30], smart city [31]–[33], and localization [34], [35], to name a few. Localization in particular, has observed a shift towards GNSS-alternatives based on cellular networks (GSM, OFDM-based) [36], Wi-Fi [37], UWB [38], etc. in recent years. While GNSS is still the leading technology for localization and real-time tracking of a subject [39] owing to advantages such as the wide coverage and availability in open space environments and high accuracy, it has equally significant disadvantages viz.,

i. high power consumption of standalone GNSS receivers.
ii. limited coverage and accuracy in harsh environments characterized by either the absence of direct LoS between the receiver and the satellite or a massive multipath phenomenon, such as urban and indoor environments [40].

A feasible solution to cope with these limitations is to reasonably integrate low-cost hardware sensors into GNSS solutions. Cooperative localization using GNSS is indeed an active field of research [41]–[43]. However, there are significant trade-offs including increased costs, complexity of localization algorithms and power consumption. Further, the rapid surge in the number of low power devices due to IoT necessitates the development of GNSS-free alternatives for localization using LPWANs. LoRaWAN is seen as a capable prospect in this context, but it is relatively new and suffers from a low adoption rate and a lack of infrastructure, especially outside of USA and Europe. Another major drawback is the lack of data pertaining to signal propagation, especially in Indian scenarios. It is well established that the propagation and performance of any wireless communication channel is affected by a number of factors such as distance, spreading factor, reflections, geography, terrains, etc. This is the reason for years of specific and extensive research in Wi-Fi [44]–[46], BLE [47], [48], 5G [49]–[51] and IEEE 802.15.4 [52], [53] despite the existence of empirical models such as the Okumura model, Hata/COST231-Hata model [54] and single-slope model [55] for radio propagation in general. Hence, an equally comprehensive look into the different aspects of communication via LoRaWAN, specifically into path loss data and propagation models is warranted to aid in link budget calculations for network planning. While a plethora of approaches based on ToA, TDoA, RSSI, AoA, etc. have already emerged, another demerit of LoRaWAN based localization is the use of GNSS at some stage, specifically in TDoA-based approaches for synchronizing time on gateways. In these cases, gateway synchronization is attained by passing the source timing information taken from GNSS (through any other technology) to the destination LoRa transceiver. This is true for certain GNSS-free deployments as well [56], only that it is used less frequently or as a one-time synchronization aid. Such an approach requires sophisticated circuitry that can cancel out timing misalignment between source and destination clocks, which comes at an increased cost. While this issue is not present in RSSI-based approaches, they have their own set of disadvantages in the form of poor accuracy, high variance in RSSI values, etc. Nevertheless, the advantages of LoRaWAN greatly outweigh these demerits, particularly with more and more countries jumping on the LoRa bandwagon. The fact that localization, in general, has become an even more integral problem in a variety of research areas such as robotics [57], WSNs [58], indoor tracking and positioning [59], and autonomous driving [60], encourages further research on LoRaWAN based solutions.

In the first half of this work, the path loss information of the LoRaWAN packets measured inside a campus building in IIIT Hyderabad is presented. The building is three-storeyed with brick walls, reinforced cement concrete ceilings, and wooden doors. The communication gateway is kept stationary, and the transmitter is placed in various locations to obtain the data. Further, the cooperation between end devices, gateways, and the network server is exploited to propose a standalone localization approach, relying only on LoRaWAN. This methodology becomes potentially attractive for GNSS-denied localization applications and is ideally meant to work with low-cost gateways. An innovative networked approach to achieve gateway synchronization without GNSS is presented, entailing an additional node in the network reserved for synchronizing the gateways. Transmissions from this node are then exploited by the gateways to simultaneously reset their timing references, resulting in synchronized ToA values. This approach also necessitates an additional synchronous counter. One potential disadvantage is that the transmissions from the additional synchronization node can easily consume a significant portion of bandwidth in an already very low-throughput network (the default LoRaWAN access scheme is pure ALOHA - based). The mitigation of bandwidth waste due to localization overhead and collisions will be tackled in future works. In this work, the feasibility of the proposed localization system is assessed as long as it is able to provide a target precision in the position estimation. This goal is accomplished by considering realistic LoRaWAN settings, incorporating duty cycle [61] limitations into the simulations.

*Chapter 2*

# Investigation of Signal Propagation and Path Loss for LoRaWAN

## 2.1 LoRa and LoRaWAN

LoRa is a proprietary modulation technique owned by Semtech. It is based on the CSS modulation technique and is specifically aimed for use in IoT applications where low energy consumption and large coverage area are the most critical requirements and high data transmission rates are relatively inessential. CSS is a spread spectrum modulation technique, where the signal bandwidth is expanded greatly beyond what is required by the underlying coded-data modulation [62]. This high bandwidth is the reason for many advantages of spread spectrum techniques (and hence LoRa) including low narrow band interference, resistance to multipath fading and jamming, robustness to channel noise to name a few.

In LoRa modulation, the spreading of the spectrum is achieved by generating a chirp signal that continuously varies in frequency [63]. The increasing (up-chirp) and decreasing (down-chirp) of the frequency is used to encode information. The rate at which this increasing / decreasing of the frequency happens is called the chirp rate. The advantage of using this method to achieve the spreading of the signal's spectrum is that the timing and frequency offsets between the transmitter and receiver are equivalent, thus reducing the design complexity at the receiver's end. Additionally, the SFs used by LoRa (defined as the logarithmic ratio of chip rate and bit rate) are inherently orthogonal, allowing the network to optimize for low power operation of the connected end nodes by adaptive variation of its power levels and data rates. The higher the SF used, the higher the range of the signal transmitted, because a higher spreading factor translates to a wider spread of the total energy of the signal, thus increasing the sensitivity of the receiver to discern a lower SNR signal. Hence, the end nodes closer to the receiver need not use high SFs, since very little link budget is needed. The link budget of any communication system can be defined in simple terms as the algebraic sum of all power losses and power gains experienced by the signal. A channel is said to be link limited when the total losses result in the incident power at the receiver to be lower than the SNR requirement. Thus, the adaptive network optimization of LoRa helps to maintain low power operation without limiting the link budget of the channel.

Although LoRa and LoRaWAN are often used interchangeably in common parlance, LoRa represents the PHY layer that enables the actual long range communication link while LoRaWAN is a networking protocol (MAC) that defines the communication protocol and system architecture for the network. These definitions have the most influence in determining the battery lifetime of a node, the network capacity, QoS, and information security [25]. While the modulation technique is proprietary, LoRa operates in unlicensed sub-GHz ISM bands, elaborated in Table 1 below.

| **Region** | **Frequency Band** |
|---|---|
| Europe | 867-869 MHz |
| North America | 902-928 MHz |
| China | 470-510 MHz |
| Korea | 920-925 MHz |
| Japan | 920-925 MHz |
| India | 865-867 MHz |

Table 1 - Frequency bands of operation of LoRaWAN in different regions.

Additionally, there are duty cycle and bandwidth restrictions in Europe and North America, with only $125/250$ kHz allowed in the former and $125/500$ kHz in the latter. The duty cycle restrictions depend on the use case, ranging from $0.1\%$ to $10\%$. These restrictions are yet to be determined in the remaining regions. The LoRaWAN network architecture is deployed in a star-of-stars topology which preserves the long range capabilities and has a high battery life for end devices when compared to the mesh network architecture used in cellular networks. The mesh network can result in a higher coverage, but because of the added complexity, it can also result in significantly higher power consumption at the end node. The star-of-stars topology essentially classifies the LoRaWAN network into three types of devices.

- The network server, that is responsible for tasks such as detecting multiple copies of packets sent by end-nodes, sending back packets to them if needed, etc.
- The gateways, that act as intermediate links between the end-nodes and the network server, relaying data from end-nodes to the network server, usually via a higher bandwidth protocol such as Wi-Fi, Ethernet, or 4G LTE and
- The end-nodes, that are the low power IoT devices that use the LoRa modulation to send information packets to gateways.

A major difference in this topology when compared to cellular networks is how the end nodes are not associated with a particular gateway [24] to have access to the network. In most scenarios, the end-nodes simply beam the packet and multiple gateways can receive it simultaneously, which are all received by the network server and it is the task of the server to distinguish between the duplicates. The end-nodes are also further divided into three classes, class A, class B, and class C based on battery lifetime and downlink latency as detailed in Fig. 1.

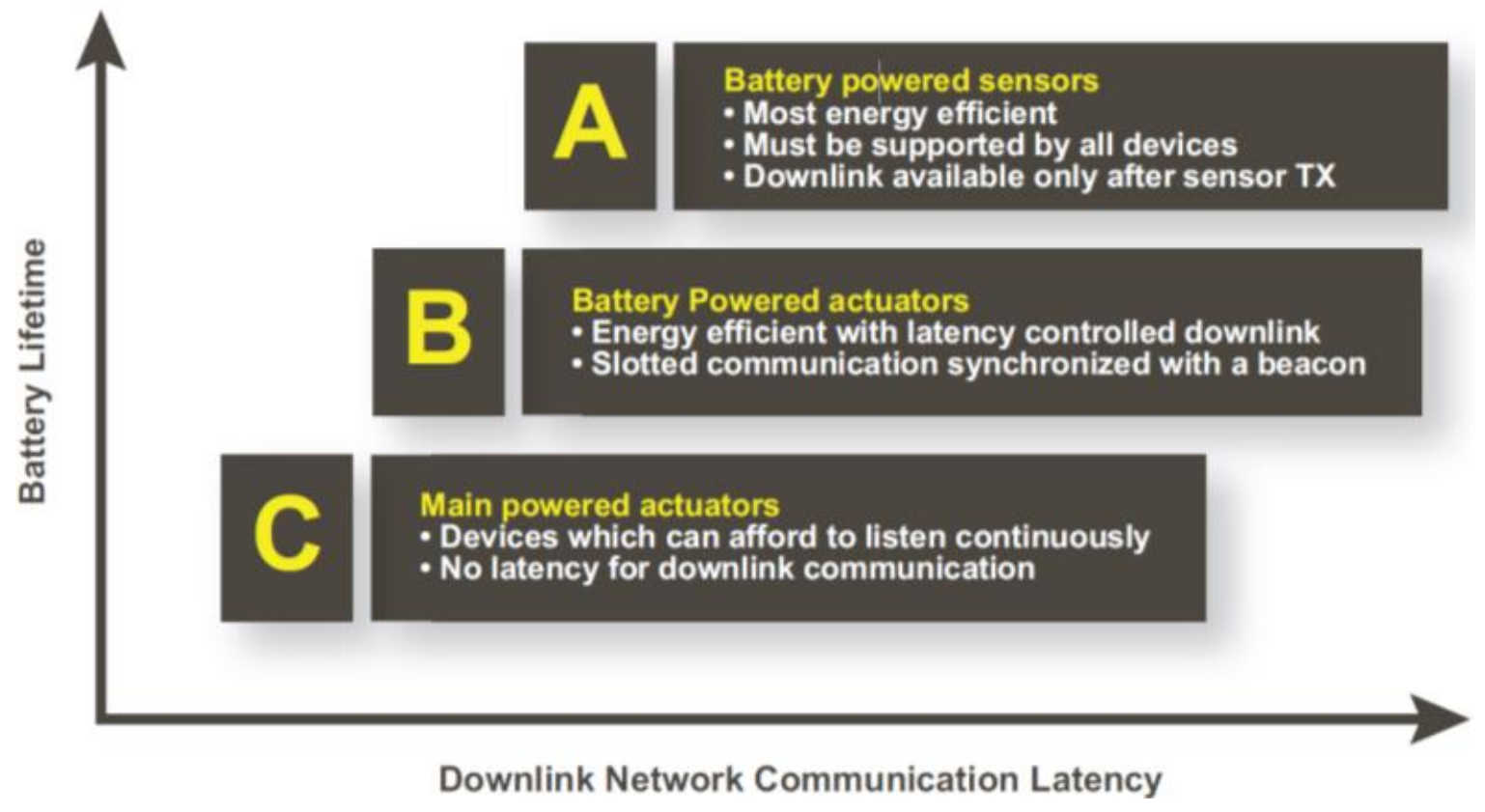

Figure 1 - Classification of end nodes in LoRaWAN [64]

All classes support bidirectional communication.

- All end nodes are required to support at least class A by default, where the communication is completely asynchronous and always initiated by the end node. Each uplink transmission is followed by two short downlink windows. End-nodes that only support class A consume the least power, because they can be immediately transitioned into low-power sleep mode when there is no data to transmit. The periodic wake-ups are also defined by the application without any network requirement. Nodes in class A communicate using the ALOHA protocol.
- End-nodes supporting class B have a deterministic downlink latency with the trade-off being higher power consumption. Periodic beacons are used to synchronize these nodes to the network, to help open downlink ping slots at scheduled times. The downlink latency can be programmed up to 128 seconds to suit the specific application. The power consumption in both class A and class B mode is low enough for the end nodes to be only battery-powered.

- Class C offers the least latency, with the obvious trade-off being the high power consumption, which also requires a continuous power source. The receiver of the end-node is kept open at all times for downlink transmission from the network server. Many devices support multiple classes of operation and also offer switching between class A and class C to suit the particular application.

## 2.2 Related Work

This section outlines the previous research carried out investigating the signal propagation and path loss of LoRaWAN. In [65], Gregora et al., present measurements of indoor propagation of LoRa signal in a reinforced concrete building in Prague, Czech Republic. The transmitter was mobile, and the receiver was stationary at the basement or on the roof in each of the two experiments. These experiments concluded that the coverage was wider when the receiver was located on the roof. The best coverage was reached in the entrance where the receiver was located, with packet loss increasing rapidly in other entrances. Petäjäjärvi et al., studied the coverage of LoRa LPWAN technology in [66], wherein a mobile node operating at 868 MHz ISM band at 14 dBm power reported data back to a base station. The maximum communication range was observed to be close to 30 km on water and 15 km on ground. as shown in Table 2 and Table 3, respectively.

| Range | Number of Transmitted Packets | Number of Received Packets | Packet Loss Ratio |
|---|---|---|---|
| $5 - 15$ km | 2998 | 2076 | 31% |
| $15 - 30$ km | 690 | 430 | 38% |
| Total | 6813 | 4506 | 34% |

Table 2 - Results of measurements with boat (Water results) [66].

| Range | Number of Transmitted Packets | Number of Received Packets | Packet Loss Ratio |
|---|---|---|---|
| $0 - 2$ km | 894 | 788 | 12% |
| $2 - 5$ km | 1215 | 1030 | 15% |
| $5 - 10$ km | 3898 | 2625 | 33% |
| $10 - 15$ km | 932 | 238 | 74% |
| Total | 6813 | 4506 | 34% |

Table 3 - Results of measurements with car (Ground results) [66].

A channel attenuation model derived from the measured data was also presented. A comprehensive study of LoRa in multi-floored buildings was presented in [67], investigating LSF characteristic, temporal fading characteristic, coverage, and energy consumption in four different buildings. It had been concluded that the one slope model could be used to predict LSF path losses in indoor environments, but only as a first-order prediction. Site-specific models were recommended for further analysis as construction materials and layout alone can change the results drastically. Temporal fading was found to follow Rician distribution, and the Rician K-factors were found to vary between 12 to 18 dB. The relationships between other parameters such as data rate, bandwidth, and packet reception rate, etc. were also studied in this work. An empirical evaluation of the indoor propagation performance of LoRa was presented in [68], where it was critically analyzed against four propagation models, ITU site generic, log-distance, multi-wall, and 3D ray tracing. It was concluded that the multi-wall model had the best performance among all four.

## 2.3 Experimental Setup

For measuring the energy loss because of various obstacles inside a building, a stationary gateway was set up as the receiver, and a mobile transmitter device was configured. The RSSI values of the packets sent from the module to the gateway were measured. The schematic plan of the building along with the location of the gateway and the locations at which the transmissions were captured are detailed in Fig. 2.

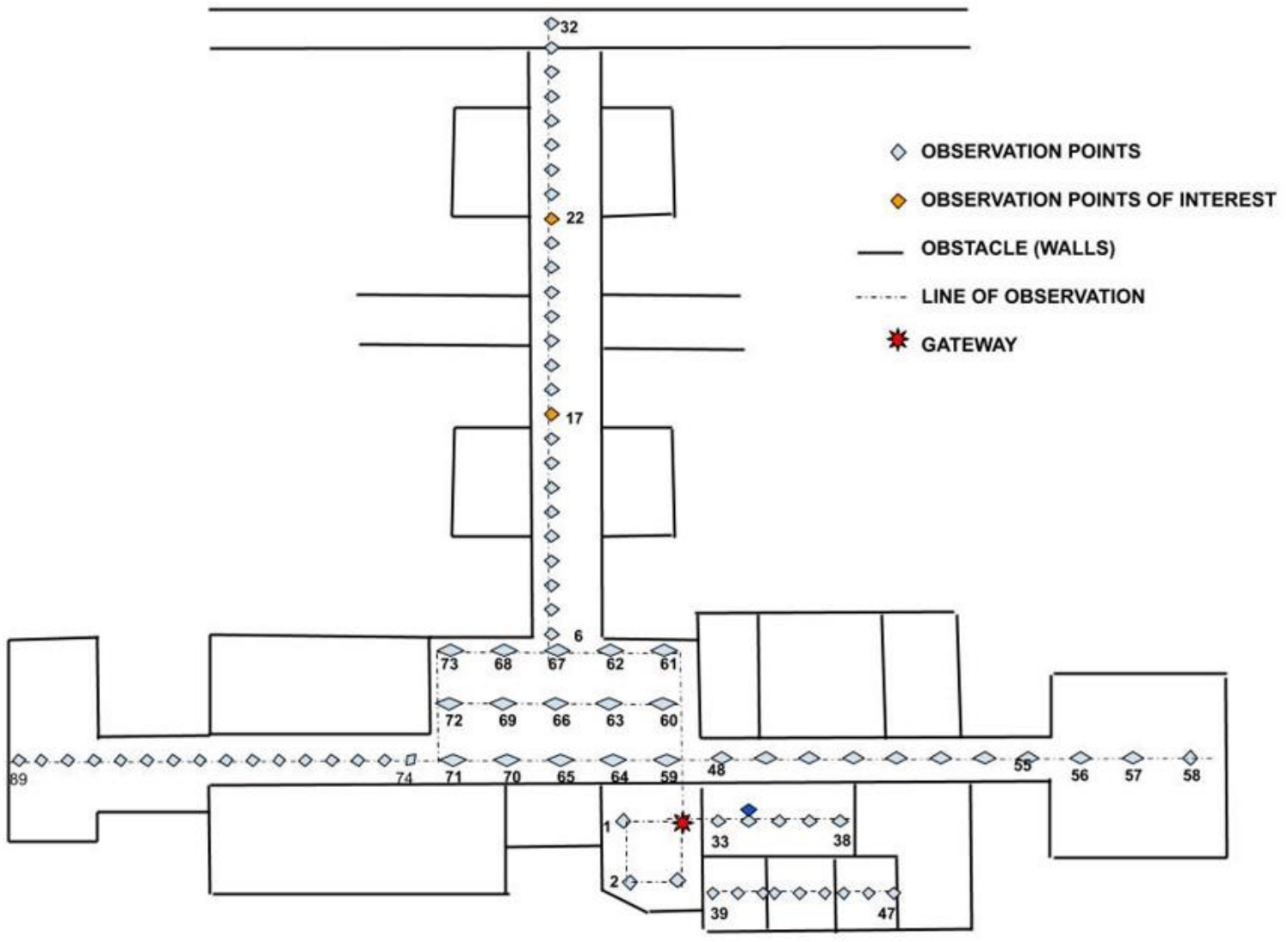

Figure 2 - Schematic plan of the building section where the experiment was carried out.

The communication is set up between a MultiConnect® Conduit™ configured as the gateway and a MultiConnect® xDot™, configured as the mobile transmitter. The xDot™ is a LoRaWAN 1.0.2 compliant, secure low power RF module, that provides long range, low bit rate, bi-directional data connectivity using sub-GHz ISM bands. The Conduit™ is a configurable communications gateway that supports dual-band Wi-Fi, Bluetooth Classic, and BLE 4.1, and has an inbuilt GNSS module for LoRaWAN packet time-stamping and geo-location capability. The MTCDT-246A-US-EU-GB gateway used for the experiment was connected to the network server through the Ethernet port (TCP/IP protocol). The gateway was set up to be stationary and the xDot™ was connected to a mobile laptop and configured to transmit dummy packets ("0000") to the gateway at regular intervals. The output power of the xDot was set to 30 dBm, the maximum transmit power for the device, and the mode of transmission was class A. The Arm® Mbed™ compatibility of the xDot™ was exploited for compiling and writing the corresponding codes. The received message at the network server consisted of the timestamp, frequency, RSSI values, SNR, data rate, etc. For the experiment, data was collected in four different sections of the building [69]. As seen in Fig. 2, the first set of locations are in the same room as the gateway to account for non-reflected, LoS signal propagation. The second set of locations from 6 to 32 are taken in the hallway outside the gateway room to account for propagation through two obstacles, a wall, and a door. The third set of locations account for signal propagation with multiple obstacles and multiple reflections, denoted by locations 33 to 47. The last set accounts for signal propagation through one obstacle, the wall between the gateway and the end-node, denoted by locations 48 to 89. Eight transmissions are sent from each of these locations and the mean and standard deviation are reported.

## 2.4 Results and Discussion

The RSSI values of the signal received from different locations in the narrow hallway (3 feet wide) i.e., locations 6 to 32 are plotted in Fig. 3. The expected result is that RSSI values should decrease with distance. While this is clearly the case as seen in Fig. 3, there are some anomalies. These are observed at distances corresponding approximately to 40 m, 80 m, and 120 m. These can be attributed to open windows in the hallway, where the route of the signal consists of multiple reflections off the buildings outside and other walls instead of passing through the two obstacles.

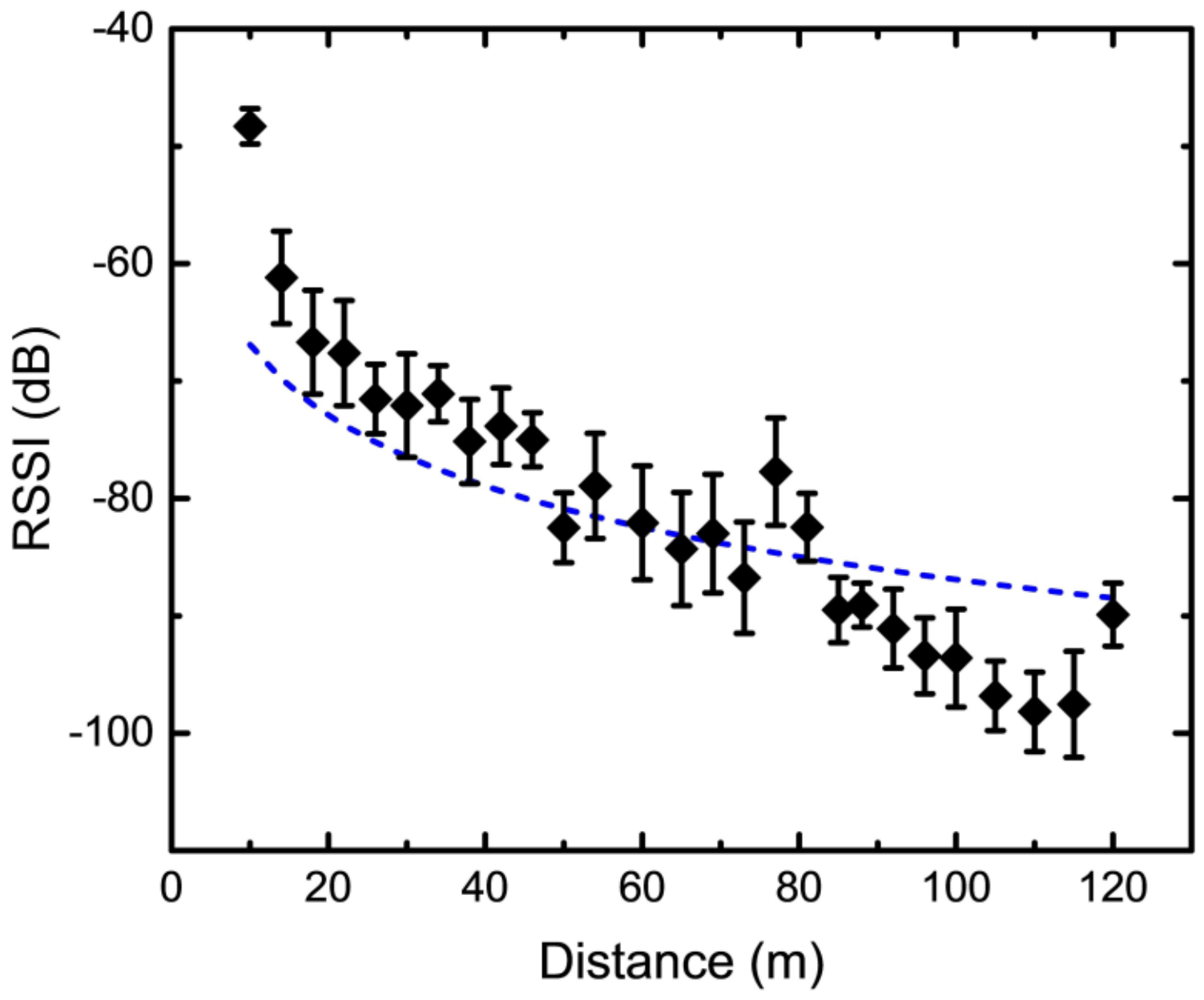

Figure 3 - Free path loss as measured inside a long narrow hallway. The black dots represent the mean values of RSSI while error bars represent the standard deviation. The blue line represents the best-fit theoretical model for LoS path loss.

The RSSI values of all the 89 locations are represented in Fig. 4. The data agrees with the expected distribution again, with the highest RSSI values in LoS conditions, followed by a single wall obstacle. It can also be seen that despite multiple walls and obstacles, the RSSI values are higher in the rooms adjacent to the gateway when compared to the long hallway, whose distance is more. This complements the expected behavior of LoRaWAN to be immune to multipath. The variation of signal strength inside each of the adjacent rooms is detailed in Fig. 5. Each room to the right adds a wall as an additional obstacle and it can be clearly seen that signal strength drops moving to the right. It is interesting to note that the drop in RSSI from room 1 to room 2 is higher when compared to others. This can again be attributed to reflections and propagation of the signal through the wooden doors instead of the brick walls in the latter rooms. This further emphasizes the complexity of modeling signal propagation in real-world scenarios.

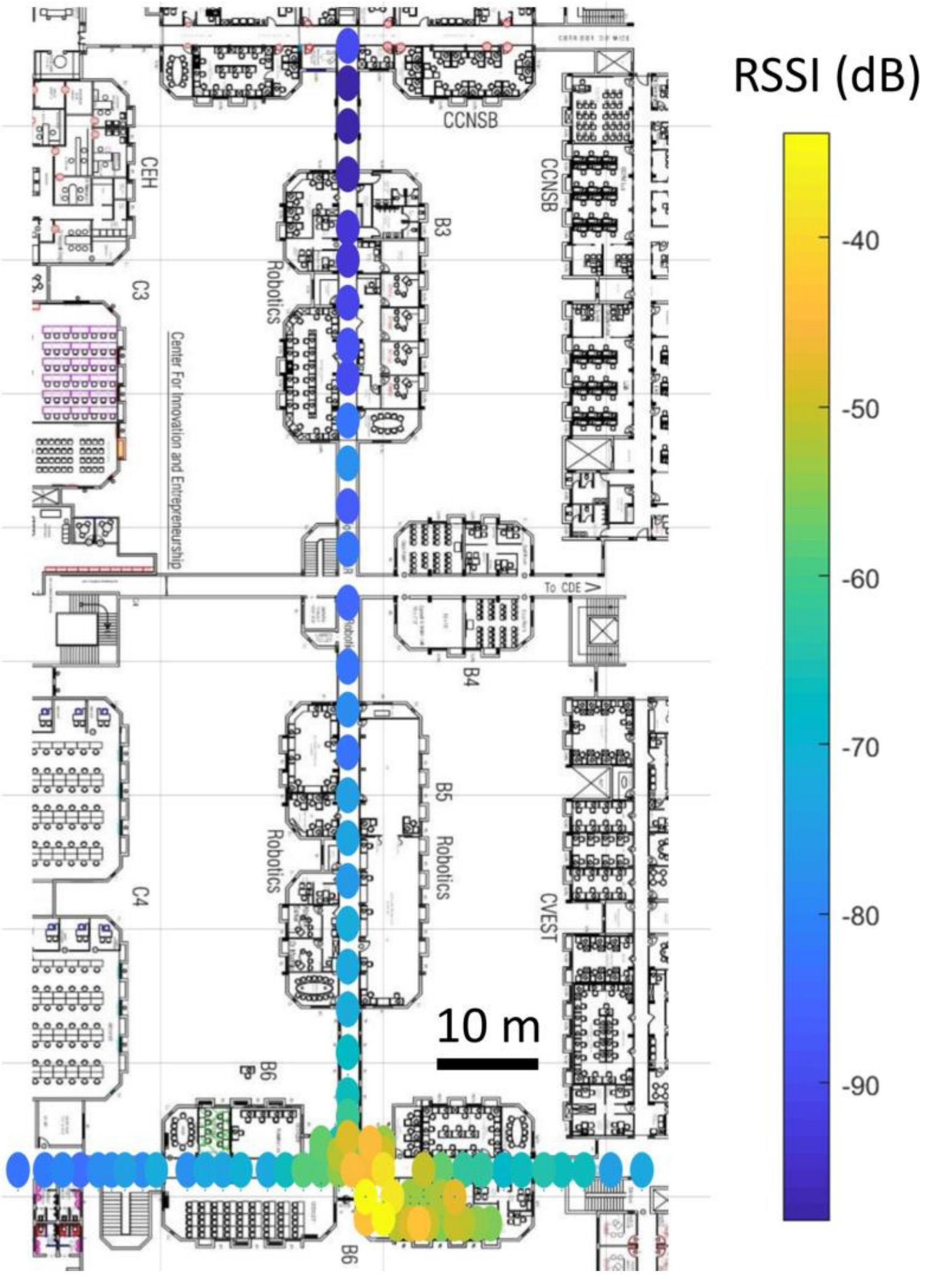

Figure 4 - Mean RSSI values for all the measured locations inside the building.

Lastly, the data outside the building at specific locations as denoted in Fig. 6 has also been presented. The RSSI values dropped from around $-90$ dB to between $-100$ dB and $-120$ dB. It has also been observed that the rate of failure of data transmissions has also increased significantly at these large distances. In this case, as well, the location of the stationary gateway is indoor, thus leading to non-LoS transmission. The RSSI ranges are $-90$ to $-100$ dB for green dots, $-100$ to $-110$ dB for blue dots and $-110$ to $-120$ dB for red dots.

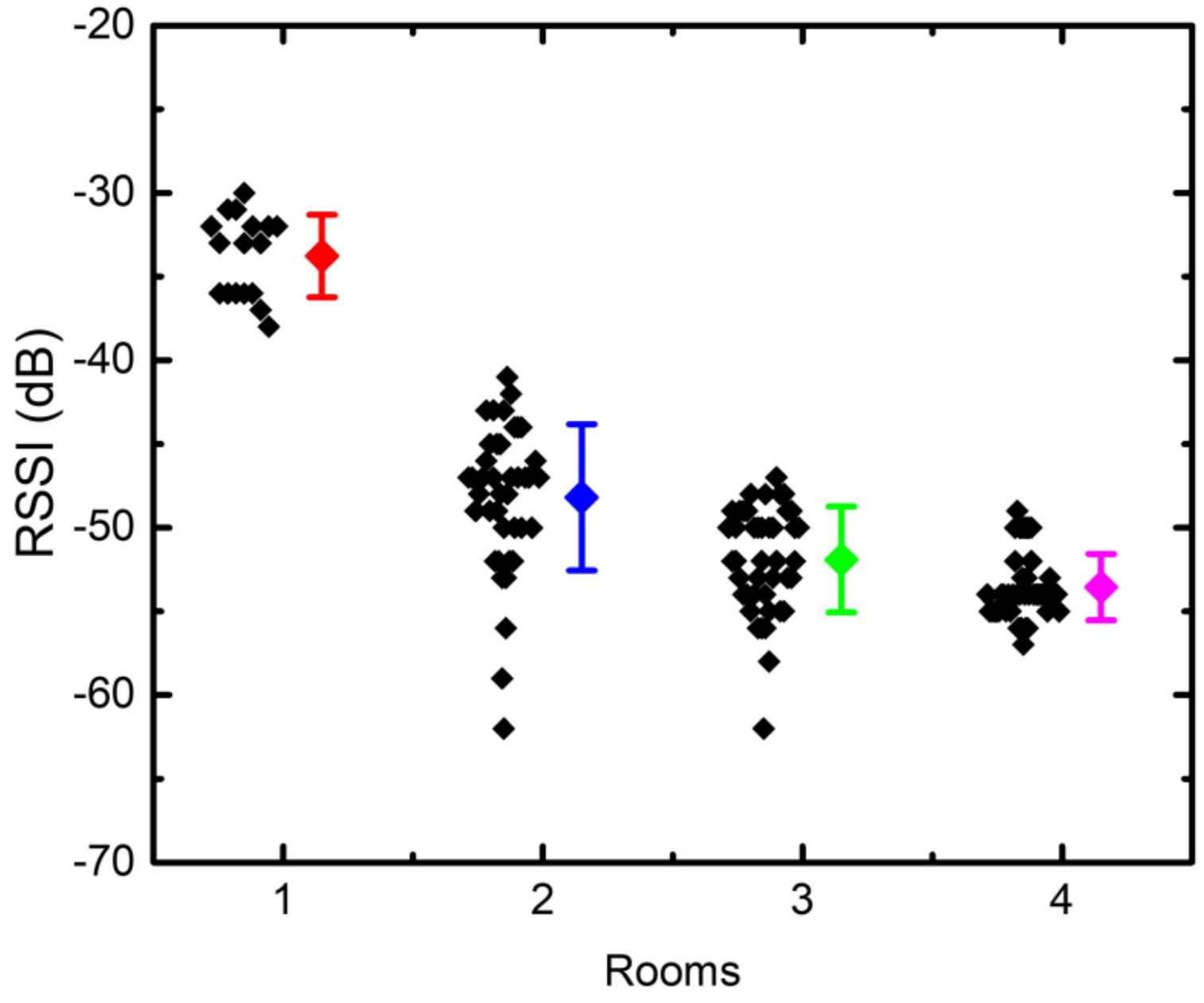

Figure 5 - Signal strength in adjacent rooms of equal size separated by a brick wall. The gateway is located in the first room.

## 2.5 Summary

The development of highly accurate path loss models for various structures is an important exercise for large-scale deployment of LoRaWAN in cities. In this chapter, the path loss information of LoRaWAN signals due to indoor obstacles such as brick walls and wooden doors inside a building has been presented. The experiments were carried out in nonideal conditions in order to consider the effect of multiple path reflections in signal propagation. Significant attenuation was observed because of the presence of obstacles such as brick walls even for very small distances. Indeed, these experiments are a first step towards obtaining critical path loss information for LoRaWAN signals in cityscapes, particularly in the Indian scenario. It is important to carry out further experiments to accurately determine the effect of the presence of multiple distributed obstacles on LoRaWAN propagation.

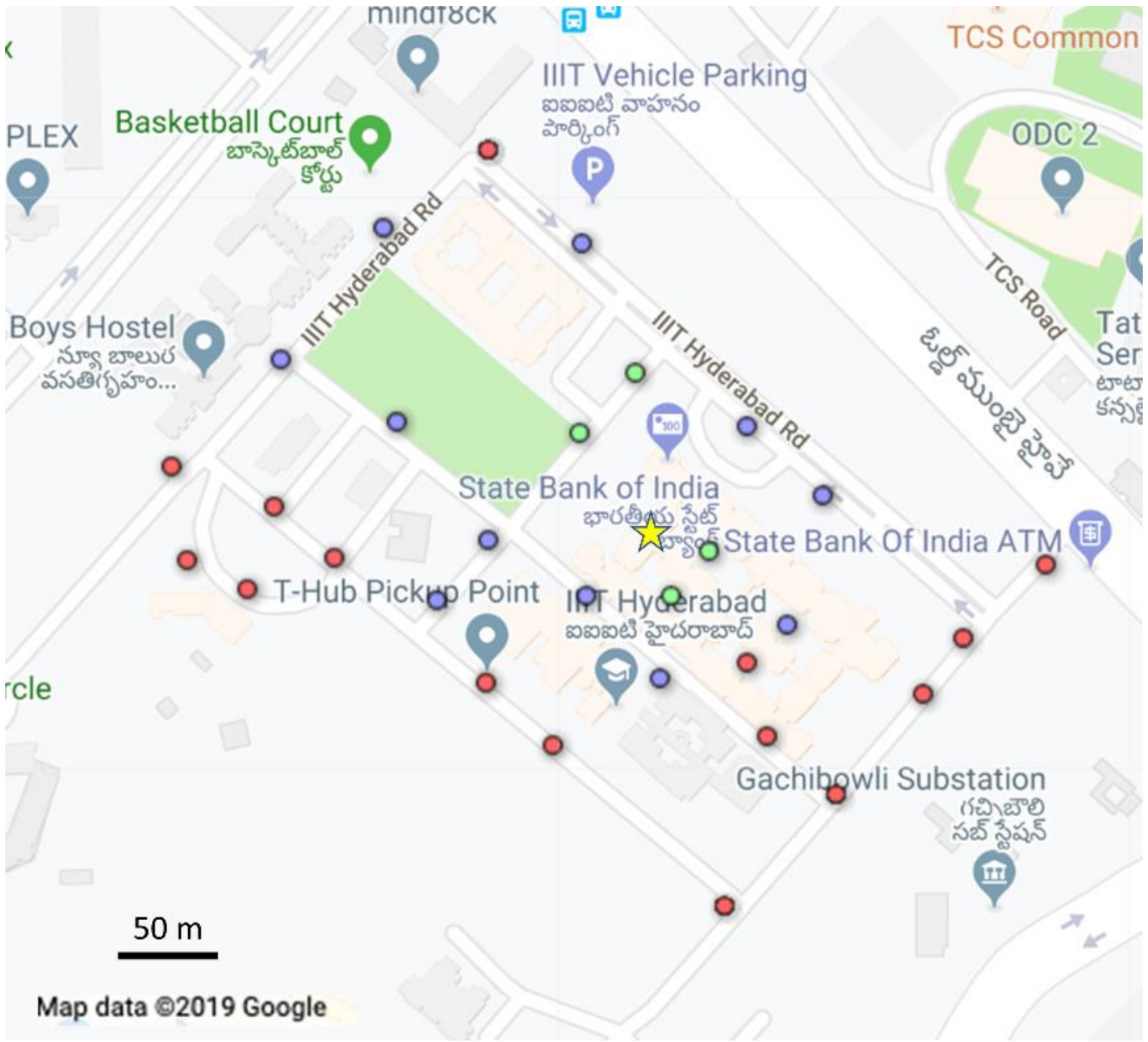

Figure 6 - Signal Strength in specific locations outside the building. The star indicates the position of the gateway.

*Chapter 3*

# GNSS-free Localization using LoRaWAN

## 3.1 Background and Related Work

As discussed already in the introduction, alternative localization technologies to GNSS have been on a steady rise. However, they are mainly plagued by poor synchronization capability between the receiver modules and the low precision of the oscillators used as timing references. Usually, these issues are tackled by the use of external modules such as GNSS timing references, or by using algorithms to achieve asynchronous localization [70]. In general, WSN-based cooperative localization is the most developed alternative to GNSS [71], using both time-based [72] and RSS-based [73] methods.

More recently, protocols such as LoRa and Sigfox emerged as potentially better alternatives, owing to much longer ranges and lower power consumption, with applications in both indoor and outdoor scenarios. In [56], Fargas et al., achieved geolocation using TDoA values obtained at LoRaWAN gateways and synchronization using GPS modules with a reported accuracy of 100 meters in locating a static device. Bakkali et al., explored Kalman-Filtering for time-based LoRaWAN localization in [74], with similarly reported accuracies.

| **Location** | **Best *k*** | **Mean Error [m]** | **Median Error [m]** |
|---|---|---|---|
| Sigfox Rural | 1 | 214.58 | 15.4 |
| Sigfox Antwerp | 10 | 688.97 | 514.83 |
| LoRaWAN Antwerp | 11 | 398.40 | 273.03 |

Table 4 – Fingerprinting Results for all LPWAN datasets. The best value of k was determined by executing a parameter sweep during the evaluation phase [75].

Aernouts et al. reported the first results of applying fingerprinting to LoRaWAN and Sigfox in [75], obtained from large rural and urban datasets with mean location errors ranging from 214.58 meters to 688.97 meters as shown in Table 4. Sadowski et al., explored RSSI-based localization techniques in [76], using four different protocols viz., Wi-Fi, BLE, Zigbee, and LoRaWAN, comparing localization accuracy and power consumption. Kwasme et al. studied RSSI-based localization exclusively using LoRaWAN in [77], concluding that frequency hopping can result in large localization errors, which can be reduced by selective averaging of single frequency RSSI readings. An experimental performance evaluation comparing externally synchronized TDoA-based and RSS-based positioning techniques in a realistic LoRa network was presented in [78], concluding that TDoA easily outperforms all investigated RSS approaches. Further studies, on a preprocessing algorithm for dropping outlier TDoA values [79] and combining AoA and TDoA measurements [80] demonstrated an improved localization accuracy. In particular, the latter demonstrated that the average error is 73% lower in comparison to the standard TDoA based approach as shown in Table 5.

| **Approach** | **Number of Estimates** | **Mean Error [m]** | **Median Error [m]** | **$95^{th}$ percentile [m]** |
|---|---|---|---|---|
| TDoA | 274 | 455 | 331 | 1167 |
| Grid-based TDoA+AoA | 319 | 286 | 221 | 657 |
| Particle Filter without AoA | 386 | 189 | 199 | 347 |
| Particle Filter with AoA | 387 | 122 | 94 | 297 |

Table 5 – Localization Results comparing standard TDoA and TDAoA [80].

## 3.2 The Proposed Idea

The principal issue with time-based techniques is the lack of synchronization between gateways and this is solved by using GNSS receivers, either inbuilt or external. To make our solution truly GNSS-free, we place a stationary synchronization node in the area covered by the gateways, whose sole task is to assure this synchronization. It is placed at a known location $(x_0, y_0)$. For example, it can be temporally "equidistant" from the gateways, i.e., a signal sent from that location will reach all gateways at the exact same time, after adjusting for the presence of real-world obstacles. This signal is a special message that specifies each gateway to reset a synchronous n-bit counter that is used to obtain the ToA values. The counters on all the gateways are reset simultaneously upon receiving this signal, providing synchronized ToA values between consecutive resets. The n-bit counter keeps time to a precision of the order $10^{-9}$ seconds.

### 3.2.1 Additional Synchronous Counter

The reason for an additional counter is that the RTC inside each gateway is not accurate enough for localization. These RTCs are also responsible for timestamping various events such as downlink and uplink transmissions, error messages, etc. which do not require high precision. Thus, they are synchronized with the outside world less frequently. However, high precision is a stringent requirement for localization; for instance, if we assume a 5 ppm drift in the RTC, the clock drift error alone accumulates to $10^{-3}$ seconds every $200$ seconds. This is completely unusable for localization as this translates to an error of the order of $10^5$ meters. To mitigate this, the synchronization with the network server needs to happen every $10^{-2}$ seconds to keep the error in the $10^{-9}$ second range (or equivalently, in the meter range). This high frequency of transmissions between the gateway and server affects the aforementioned tasks of the gateway and can also increase its power consumption significantly. Thus, a simple n-bit counter is employed, keeping time as multiples of a predetermined value. The effect of the resolution of the counter (value of $n$) on error is detailed in further sections. A noteworthy point here is the requirement of just three gateways for calculating the unknowns. This translates to a huge reduction in the setup cost, even after considering the cost of the additional synchronization node.

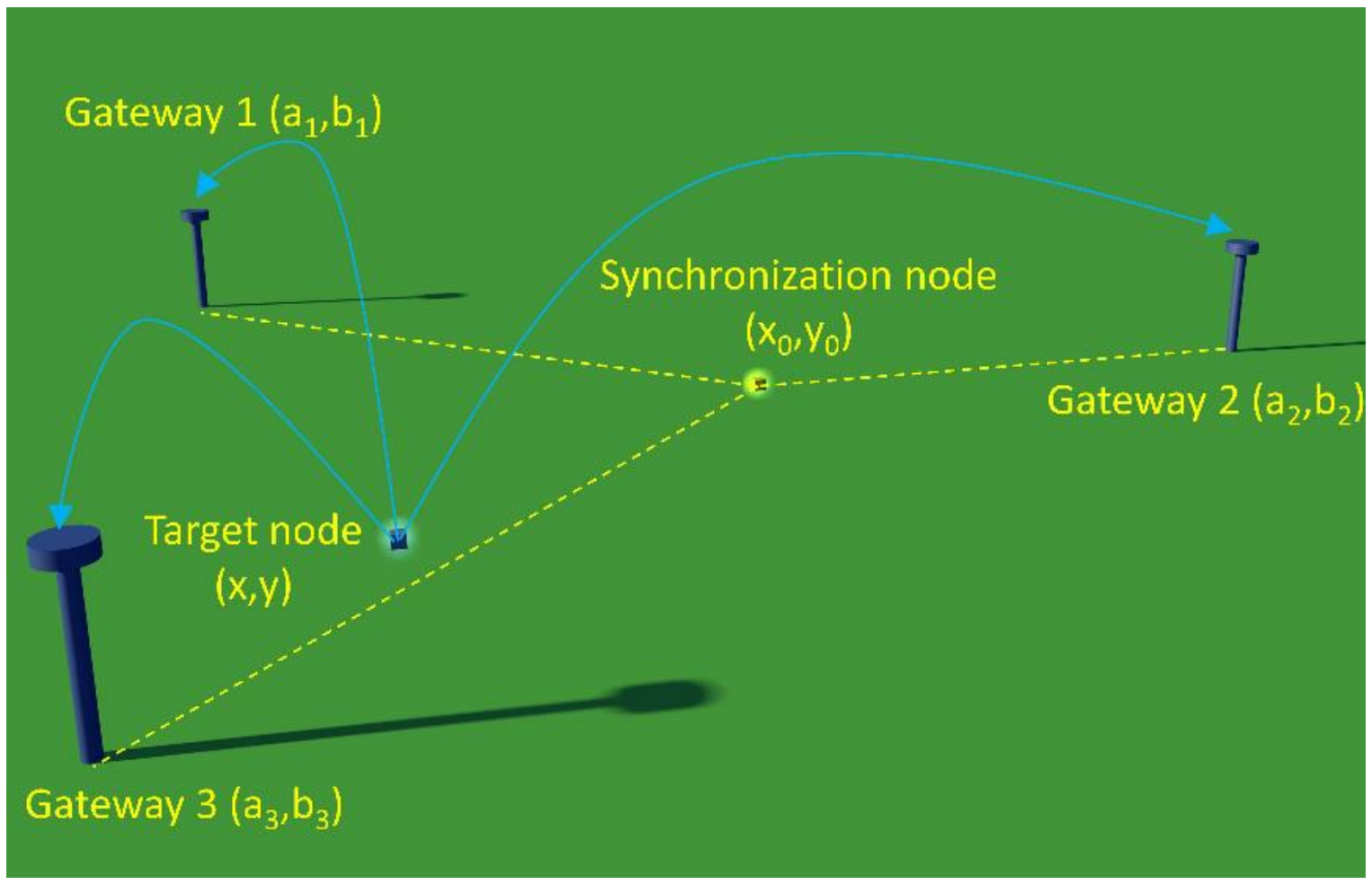

Figure 7 - A schematic representation of the proposed system. The synchronization node is placed at a known location $(x_0, y_0)$.

### 3.2.2 Position Calculation - TDoA

The location estimation of the target node after the synchronization of gateways is a fairly well-researched problem. As seen in Fig. 7, the coordinates of the three gateways are $(a_j, b_j)$ and the ToA values of the signal sent from the target node at $(x, y)$ as captured by the synchronized counter at each gateway are $t_j$, where $j$ takes integer values $1, 2$, and $3$. LoS conditions are assumed. Equating the distance between the target node and gateways to the product of the speed of the signal and time of travel gives

$$\sqrt{\left(x-a_j\right)^2+\left(y-b_j\right)^2}=c\left(t_j-t_0\right)=d_j \tag{3.1}$$

where $t_0$ is the time at which the target node initiates the transmission. The propagation speed of the signal is given by $c$, the speed of light. This nonlinear system of equations has three unknowns, $x, y$ and $t_0$. This can be solved either analytically or by eliminating $t_0$ and applying multilateration algorithms on the resulting two equations (TDoA method). One such solution is detailed in [81], as

$$x=\frac{(b_2-b_1)\gamma_1+(b_2-b_3)\gamma_2}{2[(a_2-a_3)(b_2-b_3)+(a_1-a_2)(b_2-b_3)]} \tag{3.2}$$

$$y=\frac{(b_2-b_1)\gamma_1+(b_2-b_3)\gamma_2}{2[(a_2-a_3)(b_2-b_3)+(a_1-a_2)(b_2-b_3)]} \tag{3.3}$$

where

$$\gamma_1=a_2^2-a_3^2+b_2^2-b_3^2+d_3^2-d_2^2 \tag{3.4}$$

$$\gamma_2=a_1^2-a_2^2+b_1^2-b_2^2+d_2^2-d_1^2 \tag{3.5}$$

### 3.2.3 Position Calculation – Gaussian Elimination

The three variables in equation (3.1) can be solved analytically as well, with the tricky part being the nonlinearity. It can be rewritten as

$$(x - a_j)^2 + (y - b_j)^2 + (ict_j - ict_0)^2 = 0 \tag{3.6}$$

where $j$ takes integer values between 1 and 3, and $i$ is the complex number $\sqrt{-1}$. This can be further written into vector form as

$$\|\boldsymbol{p} - \boldsymbol{w_j}\|_2^2 = 0 \tag{3.7}$$

where

- $\boldsymbol{p} = \begin{bmatrix} x \\ y \\ ict_0 \end{bmatrix}$ is the position vector of the source node, containing all the unknowns,
- $\boldsymbol{w_j} = \begin{bmatrix} a_j \\ b_j \\ ict_j \end{bmatrix}$ is the position vector of the gateways, containing all known values and
- $\|\boldsymbol{x}\|_2^2$ represents the second-order norm of the vector $\boldsymbol{x}$.

Expanding equation (3.7)

$$\boldsymbol{p}^T\boldsymbol{p} + \boldsymbol{w_j^T}\boldsymbol{w_j} - \boldsymbol{2}\boldsymbol{w_j^T}\boldsymbol{p} = 0 \tag{3.8}$$

$$\Rightarrow \quad \boldsymbol{w_j^T}\boldsymbol{p} = \frac{\boldsymbol{1}}{\boldsymbol{2}}[\boldsymbol{p}^T\boldsymbol{p} + \boldsymbol{w_j^T}\boldsymbol{w_j}] \tag{3.9}$$

Assuming

$$l = \boldsymbol{p}^T\boldsymbol{p}, \tag{3.10}$$

$$m_j = \boldsymbol{w_j^T}\boldsymbol{w_j}\ ;\ \boldsymbol{m} = [m_j], \tag{3.11}$$

$$\& \quad \boldsymbol{e} = \begin{bmatrix} 1 \\ 1 \\ 1 \end{bmatrix} \tag{3.12}$$

$l$ and $m_j$ both being scalars, and $\boldsymbol{W} = [\boldsymbol{w_j}]$, equation (3.9) results in

$$\boldsymbol{W}^T \boldsymbol{p} = \frac{1}{2}[l\boldsymbol{e} + \boldsymbol{m}] \tag{3.13}$$

$$\Rightarrow \quad \boldsymbol{p} = \frac{1}{2}[l\boldsymbol{W}^{-T}\boldsymbol{e} + \boldsymbol{W}^{-T}\boldsymbol{m}] \tag{3.14}$$

Now, we define

$$\boldsymbol{u} = \boldsymbol{W}^{-T}\boldsymbol{e} \tag{3.15}$$

$$\boldsymbol{v} = \boldsymbol{W}^{-T}\boldsymbol{m} \tag{3.16}$$

$$\Rightarrow \quad \boldsymbol{p} = \frac{1}{2}[l\boldsymbol{u} + \boldsymbol{v}] \tag{3.17}$$

Substituting equation (3.17) into equation (3.10) yields

$$l = \frac{1}{4}[l\boldsymbol{u} + \boldsymbol{v}]^T[l\boldsymbol{u} + \boldsymbol{v}] \tag{3.18}$$

$$\Rightarrow \quad (\boldsymbol{u}^T\boldsymbol{u})l^2 + (2\boldsymbol{v}^T\boldsymbol{u} - 4)l + (\boldsymbol{v}^T\boldsymbol{v}) = 0 \tag{3.19}$$

because in this case, $\boldsymbol{v}^T\boldsymbol{u} = \boldsymbol{u}^T\boldsymbol{v}$ and both are scalar quantities. Solving the quadratic equation (3.19) for $l$,

$$l = \frac{(2 - \boldsymbol{v}^T\boldsymbol{u}) \pm \sqrt{(2 - \boldsymbol{v}^T\boldsymbol{u})^{\boldsymbol{2}} - (\boldsymbol{u}^T\boldsymbol{u})(\boldsymbol{v}^T\boldsymbol{v})}}{(\boldsymbol{u}^T\boldsymbol{u})} \tag{3.20}$$

The position vector $\boldsymbol{p}$ of the target node can be calculated by substituting $l$ in equation (3.10). The computational complexity of this solution can be further improved by using orthogonal decomposition [82].

## 3.3 Error Analysis

The sources of error are the measured quantities, $t_j$ (or $d_j$, equivalently) from equation (3.1). The localization error is taken as the distance between the actual location of the target node and the calculated location. To model the errors in $t_j$, it is imperative to inspect the path of the signal (flow of packets) as shown below.

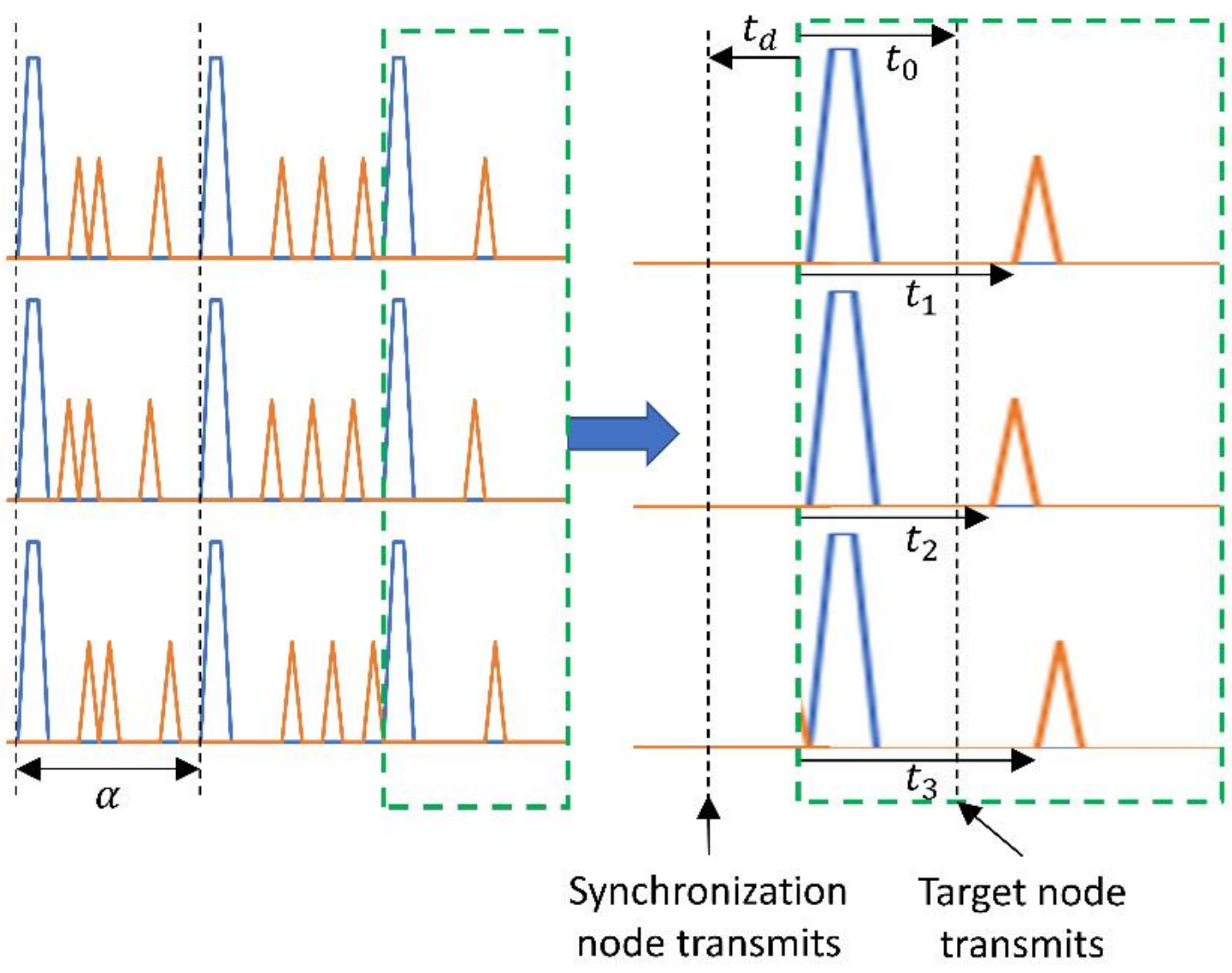

Figure 8 - An illustration of the signals received at each gateway. The blue signals are transmissions from the synchronization node, while orange represents transmissions from the target device.

The synchronization node transmits special packets every $\alpha$ seconds, denoted by the blue pulses in Fig. 8, which is a simplistic representation of the dataflow focusing on the signal path, ignoring channels and bands. Building on our assumption of temporal equidistance, Fig. 8 depicts the blue pulses to be arriving at the same instant at all gateways, i.e., $t_{d_j} = t_d$ for all three gateways, where $j$ takes integer values between 1 and 3, and $t_{d_j}$ is the ToA value of synchronization node transmission at the $j^{th}$ gateway. However, this is not compulsory. The values of $t_{d_j}$ can be different for each gateway and still aid in synchronization, as they can always be calculated from the known position of the synchronization node. This perpetual knowledge of the synchronization node's position is the only requisite. The choice of $\alpha$ depends on the resolution of the counter used, which itself depends on a slew of factors such as power consumption, design complications, the cost of the setup, etc. The orange triangular pulses in Fig. 8 denote the packets sent from the target node. For the $t_j$ values to be accurate, orange pulses/target node transmissions should only occur between consecutive synchronization node transmissions/blue pulses. In reality, the time gap between the counter overflowing and resetting can be neglected relative to the on-time of the system. Hence, this system can essentially stay "always-active", with almost zero downtime.

Also, the resetting behavior of the counter means that $t_j$ is non-negative for any gateway. $t_0$ is also non-negative because it is always measured with respect to the previous synchronizing signal, and $t_{d_j}$ is always negative as the transmission from the synchronization node always happens before the synchronization. The possible sources of error in the above-outlined data path are

i. Error in the position of synchronization node,
ii. Oscillator drift in the crystals of both the additional counter and inbuilt processor of the gateway,
iii. Least count of the counter (rounding off error) and
iv. Clock slippages in the gateway processor during various operations. For instance, there will be a delay between receiving the synchronization packet and the counter resetting. A similar analogy can be drawn to other transmissions and operations, such as calculating the ToA values. There will be a delay between receiving the target node packet and the counter registering the ToA value. This delay can be significant in this scenario because time is being counted in the order of $10^{-9}$ seconds.

Modeling the error in $t_j$ with these variables [83],

$$e_{t_j} = \Delta t_{d_j} + N_j \omega_{1j} + \mathcal{U}_{2j}[0, T] + \mathcal{U}_{1j}[0, k_j] \cdot (T_g + \omega_{2j}) \tag{3.21}$$

where $j$ represents the $j^{th}$ gateway. The Gaussian distributions $\omega_{1j} \sim \mathrm{N}(0, \sigma_{1j}^2)$ and $\omega_{2j} \sim \mathrm{N}(0, \sigma_{2j}^2)$ with zero mean and adjusted variances model the drift in the counter and processor clock, whose time-periods are denoted by $T$ (frequency = $f$) and $T_g$ (frequency = $f_g$) respectively. $N_j$ is the multiple of $T$ stored in the counter corresponding to the $t_j$ value, that varies between 0 and $2^n$. $\mathcal{U}_{1j}$ is a discrete uniform distribution to model processor clock slippages ($k_j$ being the maximum number of clock slippages) and $\mathcal{U}_{2j}$ is a uniform distribution to model the rounding-off error in the $n$-bit counter. $\Delta t_{d_j}$ denotes the offset due to the error in the position of the synchronization node, given by

$$\Delta t_{d_j} = \frac{(x_0 - a_j)\Delta x_0 + (y_0 - b_j)\Delta y_0}{c^2 t_{d_j}} \tag{3.22}$$

Equation (3.21) is the comprehensive model of total error possible in this scenario whereas, in reality, there might not be two different crystal oscillators for the processor and the $n$-bit counter. For instance, the clock of the counter can be generated from the main processor clock using frequency multipliers. To determine the critical design parameters of this system, it is important to consider the relative contributions of these components in the total error. In equation (3.22), the distance between the synchronization node and the $j^{th}$ gateway is given by $ct_{d_j}$, which is of the order of $10^3$ to $10^4$ meters. The difference in coordinates in the numerator is also of the same order. Errors in synchronization node coordinates denoted by $\Delta x_0$ and $\Delta y_0$ respectively, are ideally in the $10^{-2}$ to $10^{-1}$ meter range, which implies that $t_{d_j}$ is in the range of $10^{-9}$ to $10^{-10}$ seconds. If we assume the processor frequency to be 400 MHz ($T_g$ = 2.5 ns) which is the same as the ARM9 Processor found in the MultiConnect® Conduit™ (Model Number: MTCDT- 246A-US-EU-GB) used in the previous chapter, an exaggerated 100 clock slippages will still result in an error of the order of $10^{-7}$ seconds. Similarly, if the counter clock frequency is assumed to vary between 10 MHz and 200 MHz, the rounding-off error is of the order of $10^{-7}$ seconds. As stated already in subsection 3.2.1, clock drift error can be in the range of $10^{-3}$ to $10^{-2}$ seconds, which implies it is indeed the major contributor.

However, clock drift is an extensively researched problem at the same time. Compensation techniques range from the use of DTDoA [84] and novel time synchronization schemes [85], [86] as after-the-fact solutions to using techniques such as temperature compensation while designing the oscillator itself [87], [88]. Also, none of the experimental results detailed in subsection 3.1 discuss clock drift errors, further reinforcing our assumption that they can be predicted and canceled out. Hence, we assume that drift is negligible and discard it from our model in equation (3.21). Another important constraint to consider in our pursuit of optimizing for the least error is the duty cycle ($\delta$), which is restricted for LoRaWAN in some geographies as already stated in section 2.1. Thus, we need to conduct this error analysis from a design problem perspective.

## 3.4 Design Problem

The goal of modeling errors and running simulations is to optimize for the least error in localization while staying within the duty cycle limitations. Hence, the maximum acceptable mean error in localization $(e_{max})$ and maximum duty cycle allowed $(\delta_{max})$ are our primary design constraints. For this section, we assume the clock slippages and drift to be negligible, thus, equation (3.21) becomes

$$e_{t_{ideal}} = \mathcal{U}_2[0, T] \tag{3.23}$$

Or

$$e_{t_{ideal,max}} = T \tag{3.24}$$

The Duty cycle is the ratio of the time-on-air $(\tau)$ to the total time-period between successive transmissions $(2^n T)$. This ratio should always be less than the maximum allowed duty cycle. Thus,

$$\delta = \frac{\tau}{2^n T} \leq \delta_{max} \tag{3.25}$$

which implies that the primary design variables are $\tau$ , $T$ (or $f$), and $n$. T is determined by $e_{max}$, which is then used to choose the appropriate values of $\tau$ and $n$ to satisfy the constraint of $\delta_{max}$. However, $e_{max}$ can be simulated only when the orientation of gateways is known. This depends on the range of LoRa in the deployment terrain, which depends on the SF used [27]. Hence, SF acts as a secondary design constraint, limiting the range $\tau$ of before duty cycle constraints and affecting the values of $T$ indirectly. Time-on-air is also dictated by the SF used, among other factors such as symbol duration, bandwidth, data rate, etc. To further understand the effects of the above parameters on time-on-air, the format of the LoRa packet needs to be studied.

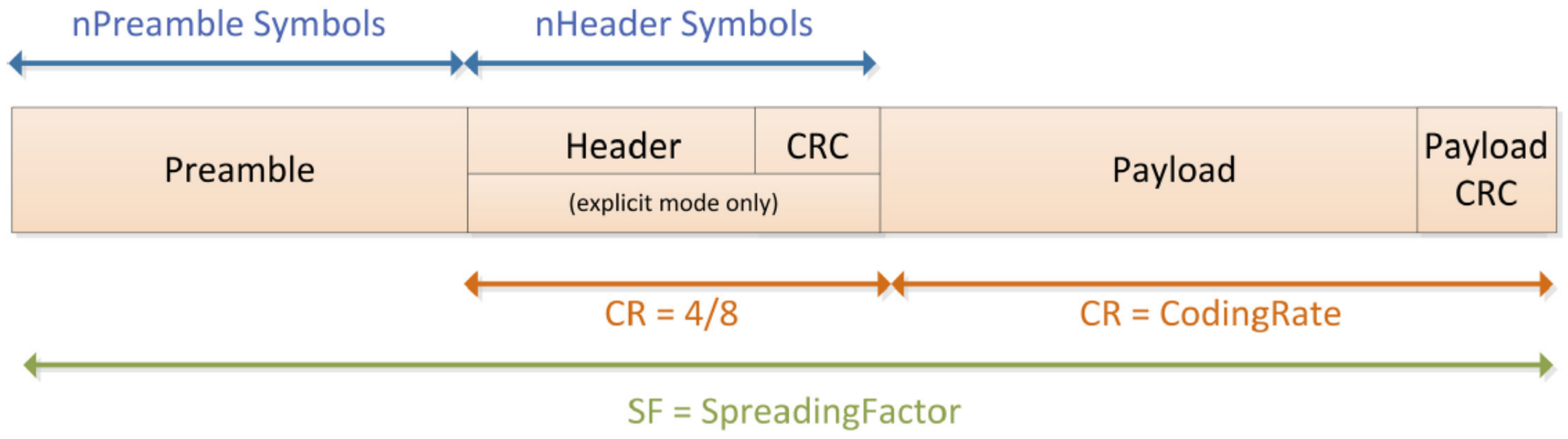

Figure 9 – LoRa Modem Packet Formatting [89]

As seen in Fig. 9, the LoRa packet consists of several elements including a common preamble for all modem configurations, a header element if the packet length is not known in advance (which is usually the case), a payload element that consists of the information that needs to be relayed and CRC frames. If we define the symbol duration as $T_{sym}$, then

$$T_{sym} = \frac{2^{SF}}{BW} \tag{3.26}$$

where BW is the bandwidth of the signal that also defines the chip rate.

The duration of the preamble is given by [89]

$$T_{preamble} = \left(n_{preamble} + 4.25\right) \cdot T_{sym} \tag{3.27}$$

where $n_{preamble}$ is the number of programmed preamble symbols.

The total number of symbols in the payload and header are given by [89]

$$n_{payload+header} = 8 + max\left(ceil\left(\frac{8PL - 4SF + 28 + 16 - 20H}{4(SF - 2DE)}\right) \cdot (CR + 4), 0\right) \tag{3.28}$$

where

- $PL =$ number of bytes in the payload
- $H = 0$ when the header is present and $H = 1$ when no header is present (if the packet length is known in advance)

- $DE = 1$ when low data rate optimization is enabled and $DE = 0$ when it is disabled
- $CR$ varies from 1 to 4.

As seen in equation (3.28), if the packet length is known in advance, the header can be removed to reduce time-on-air. With the number of bytes in the non-preamble part of the packet defined, payload duration is simply

$$T_{payload} = n_{payload+header} \cdot T_{sym} \tag{3.29}$$

and time-on-air, $\tau$ is

$$\tau = T_{payload} + T_{preamble} \tag{3.30}$$

Another secondary design constraint is $T_g$, limiting the precision to choose $T$. As stated already, the additional clock for the n-bit counter is likely to be generated from the processor clock, hence a faster processor clock or lower $T_g$ gives us higher precision to choose $T$. Lower $T_g$ also reduces error due to clock slippages as seen from equation (3.21). In the set of values of $T$ allowed by $e_{max}$, the maximum value should be chosen to minimize the duty cycle.

In the case of $n$, a lower bit resolution drastically reduces the effect of oscillator drift but increases the duty cycle exponentially at the same time. Hence, it needs to be tuned accordingly. Similarly, $\tau$ also needs to be tuned to minimize duty cycle by varying SF (if the LoRa coverage allows multiple SFs in that particular deployment), coding rate, etc. $T$ and $n$ are further constrained by the overflow of the counter; faster clock and low n result in a quicker overflow, while slower clock reduces the precision of the measured time. Hence, the values for the variables $\tau$, $T$ and $n$ have constraints and both upper and lower bounds, requiring the system design to be a trade-off between these depending on the deployment environment.

## 3.5 Simulations and Results

To simulate for the maximum error possible with our assumptions, the gateways are positioned at the vertices of an equilateral triangle circumscribed by a circle of diameter 10 km and centered at the origin, effectively resulting in a maximum transmission distance to be around 8.6 km (the side of the triangle). This coverage is achievable only by using the highest spreading factor, SF12. While higher ranges up to 15 km are possible on ground deployments as evaluated in [66], the same study also calculated packet losses to be as high as 74% beyond 10 km. This can prove to be costly in the context of localization, especially in the case of synchronization packets in our architecture. The vertices of the triangle of gateways are at $(0, -5000)$, $(4330.13, 2500)$, and $(-4330.13, 2500)$. The location of the target node is generated randomly inside this triangle, and the corresponding TDoA values are calculated from equation (3.1). The processor clock is assumed to be 400 MHz ($T_g$ = 2.5 ns). The first simulation establishes the relation between $e_{max}$ and T. For the calculation of $e_{max}$, 106 points are taken at random inside the triangle. $e_{t_{ideal}}$ is taken to be its maximum value as given in equation (3.24) and to further assure that the maximum possible value of $e_{max}$ is calculated, the 8 permutations that arise due to the sign of $e_{t_{ideal}}$ when added to the TDoA values corresponding to each of the 106 locations, are also considered. Thus, each of the 106 locations has 8 erroneous estimates calculated from the 8 erroneous sets of TDoA values as shown in equations (3.2) to (3.5). The maximum among those values is taken as localization error for that point. The mean of all the 106 values is taken to be $e_{max}$. The value of $T$ is varied between 2.5 ns and 100 ns in steps of 2.5 ns and for each value of T, the same 106 points are considered and $e_{max}$ is subsequently calculated. The results are shown in Fig. 10.

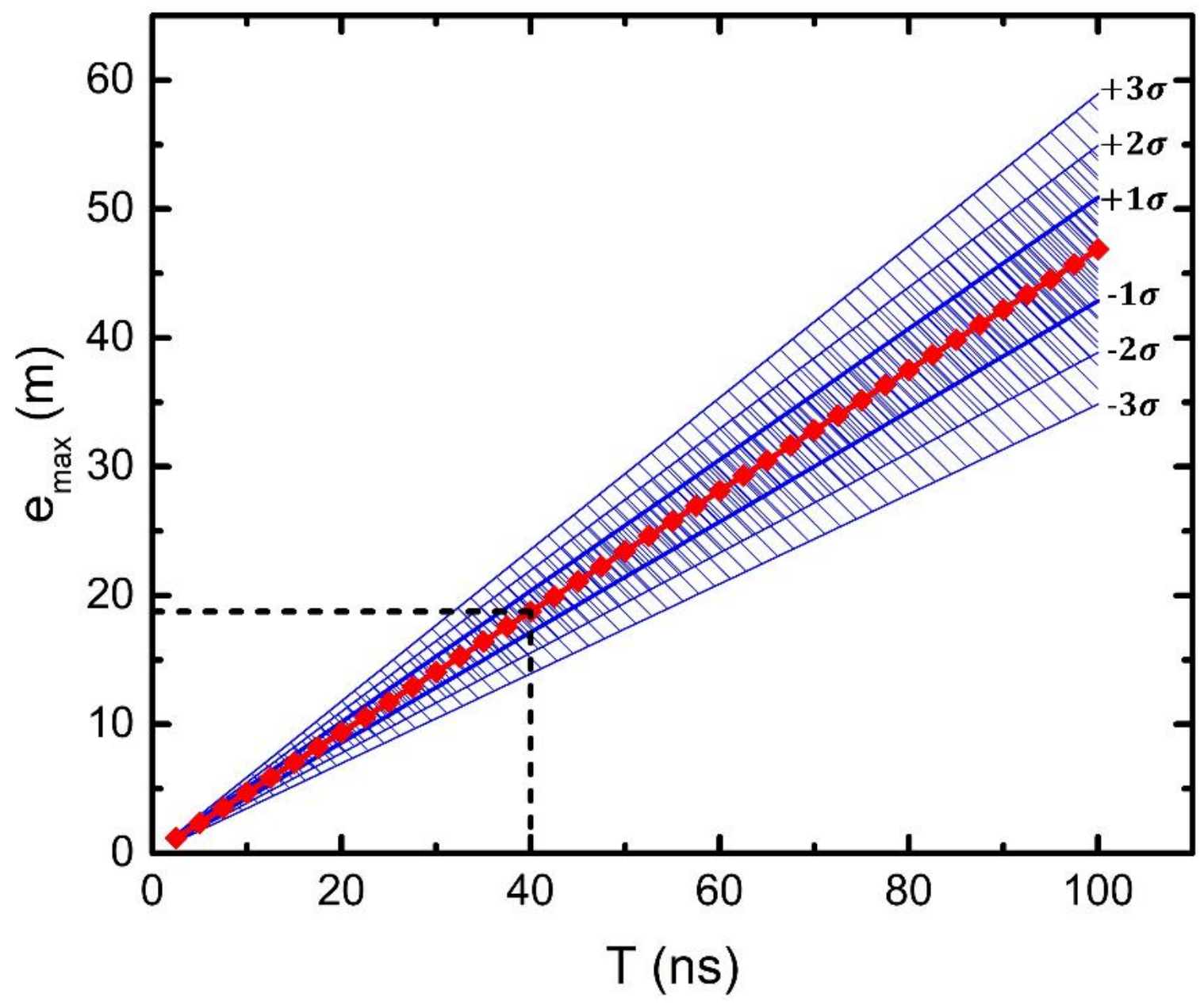

Figure 10 - Variation of $e_{max}$ with $T$. $e_{max}$ is the maximum acceptable mean error at any position within the triangle and T is the time period of the n-bit counter clock. The shaded regions indicate the confidence intervals of $1, 2$ and $3$ sigmas.

As stated already, the value of $e_{max}$ limits $T$. Essentially, this translates to an upper bound on the time-period of the counter clock $T$, denoted by $T_{max}$ or a lower bound on its frequency $f$, denoted by $f_{min}$. The actual value to be taken is then determined by other considerations, duty cycle being the foremost among them. The next simulation aims to reveal the restrictions because of the duty cycle constraint. By making $T$ constant in equation (3.25), the choices of and $n$ are explored. Since we are using only SF12, the values of α are restricted between $0.25 - 3.60$ seconds. These bounds have been calculated with the aid of open-source air-time calculators available online, considering all possible bandwidths ($125, 250$, and $500$ kHz), load packet lengths ($51$ bytes or lower for $125$ kHz, $33$ bytes or lower for $500$ kHz), and coding rates $(0.5 - 0.8)$. We assumed $T = 40\ ns$ ($f = 25$ MHz). This choice assures a reasonably good precision while also keeping $e_{max}$ at $18.75$ m (Fig. 9). For a $32$-bit counter, this also translates to an overflow time of $171.79$ s, which is again a good balance. A key point to note here is the lack of flexibility to choose $n$, even for a duty cycle as high as $10\%$ as seen in Fig. 11a. While lower values of $n$ like $4$-bit and $6$-bit counters for instance, drastically reduce drift as seen in equation (3.21), it has an equally catastrophic effect on the duty cycle because of the need for repeated transmissions. However, in the presence of drift compensation, it is reasonable to choose a $32$-bit counter to reduce the duty cycle of the synchronization node. Lower resolutions, such as a $28$-bit counter, for instance, are viable options only when the duty cycle is not very important and there are significant gains in other parameters such as clock drift, low power consumption, simpler design, etc. It should be noted that the duty cycle variation in Fig. 11 is only for a predetermined constant value of $T$ stated above and a slower clock can also be considered if the duty cycle is of higher importance.

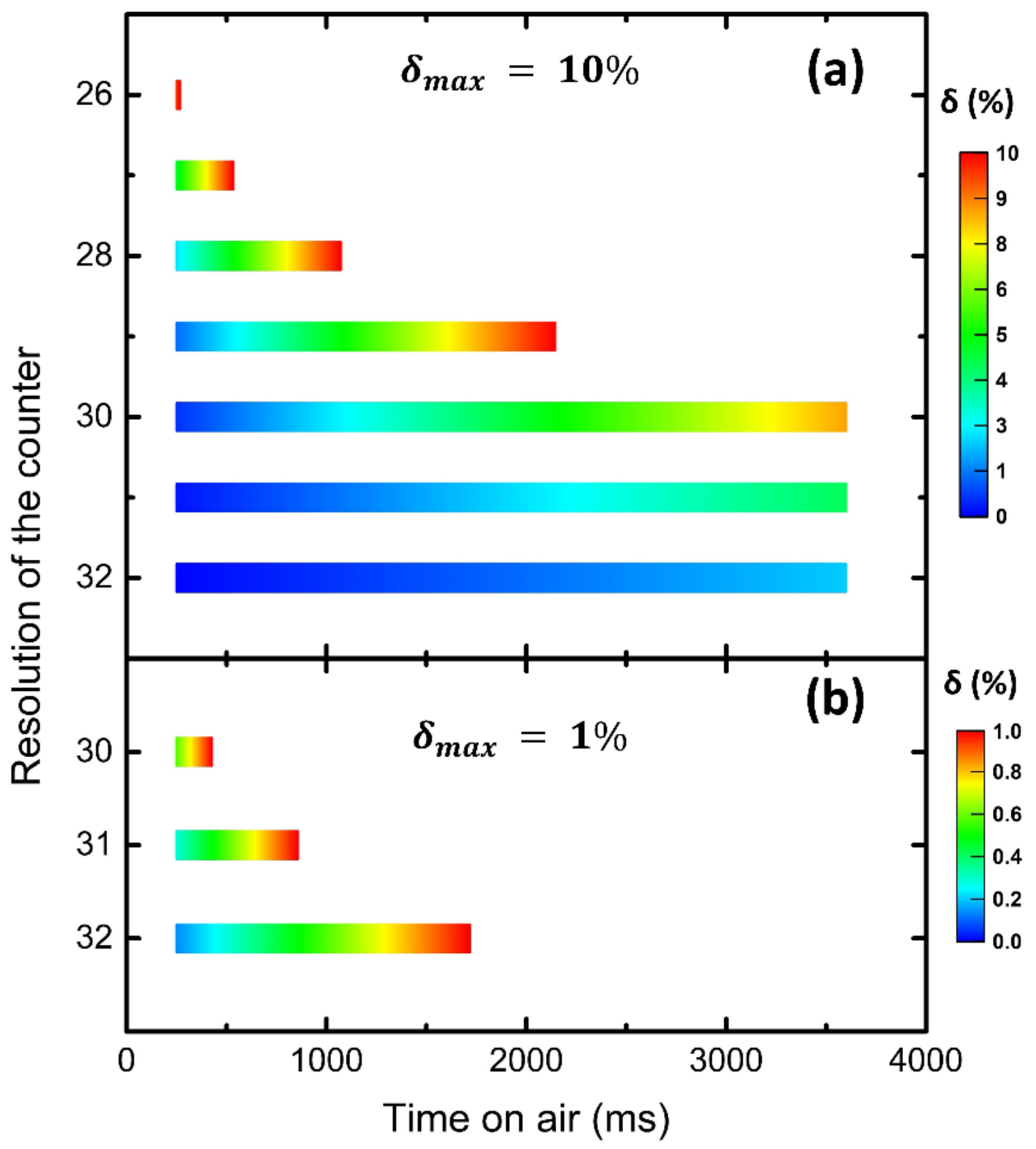

Figure 11 - Variation of duty cycle with the time-on-air $(\tau)$ and the resolution of the counter $(n)$. (a) limited to 10%. (b) limited to 1%. The value of $T$ is taken as 40 ns.

Finally, the variation of localization error within the triangle of gateways is calculated based on the values of the design parameters as $T = 40\, ns$ and $n = 32$ to serve as a basis for future studies. $7050$ points inside the triangle are taken as the locations of the target node with et for each corresponding set of TDoA values given by equation (3.23). $23$ values of $e_t$ are taken at each location to indicate $23$ transmissions from each position of the target node, each time with a different error in the values of $t_j$ to mimic a real-world scenario. The same $8$ permutations considered in Fig. 10 are also considered, and the maximum value of localization error among those $8$ permutations and $23$ transmissions is taken as the error at that location.

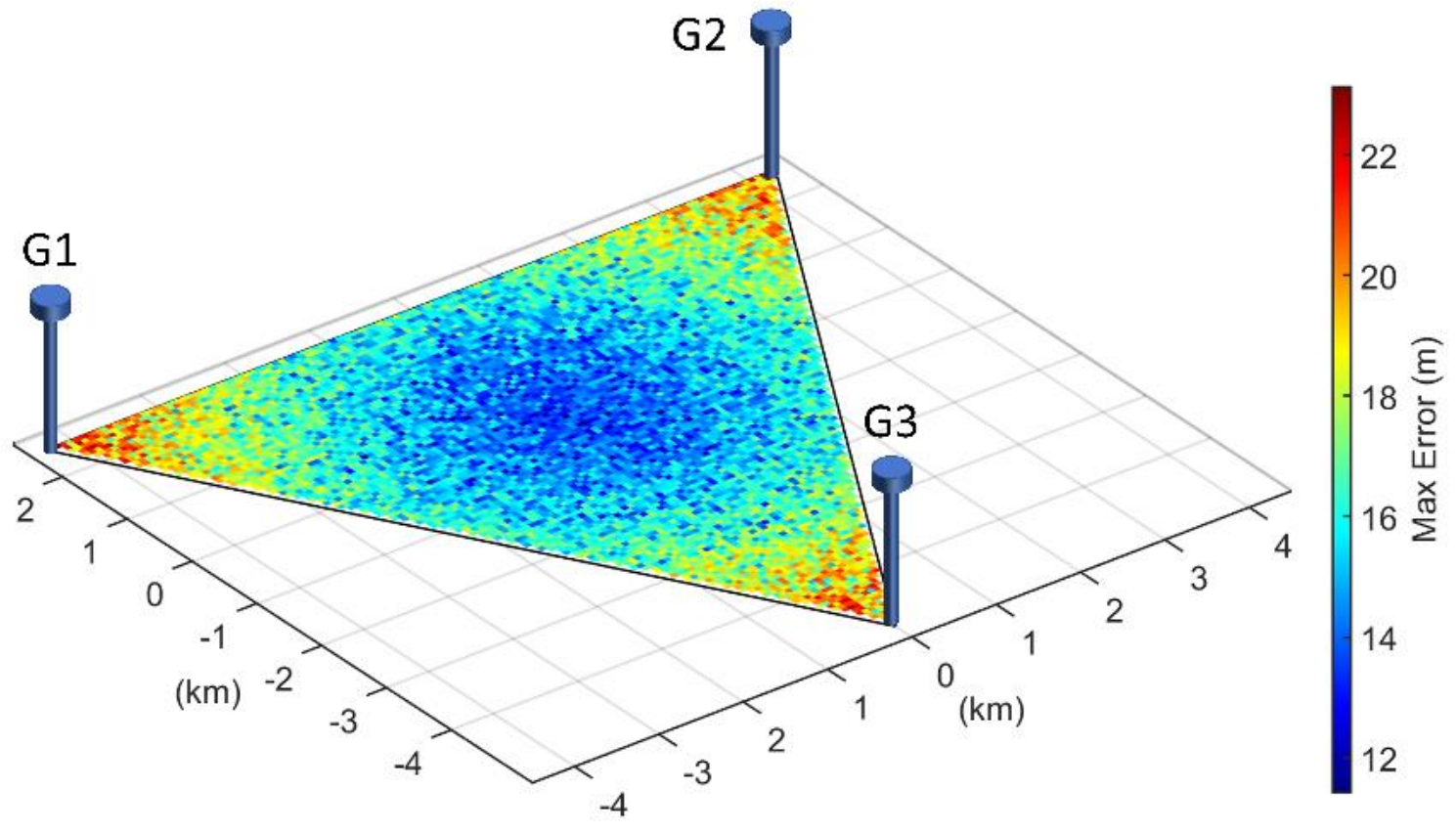

Figure 12 - Distribution of maximum localization error within the triangle for $T = 40$ ns, and $n = 32$.

The distribution of this localization error is as shown in Fig. 12. The maximum error obtained in this simulation is approximately $23$ meters. This is higher than the estimated $18.75$ meters from Fig. 10 because that is an estimate of the mean. The distribution in Fig. 12 shows the higher error is likely closer to the gateways (and away from the synchronization node), compared to the center of the triangle. This is intuitive because all the time of arrival values are relatively large leading to less effect of errors, compared with values from the edge of the triangle, where one of the values is very low.

*Chapter 4*

# Conclusion and Future Work

In this work, the prospect of standalone LoRaWAN based localization, specifically for IoT applications has been investigated. One of the major demerits of LoRaWAN deployments in India has been addressed by investigating the signal propagation and path loss of LoRa packets in an urban environment. This has been achieved by experimental modeling of the two parameters in a campus building inside IIIT Hyderabad, drawing some interesting conclusions about the multipath resistance of LoRa. This can be further extended to a campus-wide and region-wide performance evaluation of LoRaWAN signal propagation and path loss, and such evaluations are mandatory as well before the deployment of a LoRaWAN based IoT network.

A novel and standalone TDoA-based localization approach using LoRaWAN was proposed. The problem of synchronization between the gateways was solved by a GNSS-free, collaborative methodology entailing a stationary synchronization node. An error analysis for the approach has also been presented detailing the effects of many factors such as drift, clock slippages, and clock frequencies on the localization error. The consequences of duty cycle restrictions were also studied, modeling the analysis as a design problem with localization error and duty cycle limitation as design constraints. Finally, simulations exploring the relationship between design variables such as resolution and clock frequency of counter and time-on-air with the design constraints were also carried out. The distribution of localization error at different locations inside the triangle of gateways was also simulated. The extent of future work in this context is vast, as this work largely deals with proposing the initial idea and theoretical considerations for the setup.

The reason for such a distribution with low errors in the central region of the triangle and the effect of adding clock drift and slippages are some of the interesting insights that we are exploring in more detail. Furthermore, the impact of both synchronization signal overhead and frame collisions on the bandwidth usage will be carefully investigated in order to bound the range of applications benefiting from such an approach. Remarkably, the synchronizing function of the synchronization signals can also be performed by frames sent from LoRa-enabled smartphones carrying GNSS information. This possibility introduces an architectural simplification (synchronization node not required) but raises a wide spectrum of research challenges mainly due to the unpredictable position of the synchronizing devices and the consequent impact on network performances. We are also currently working on extensive experimental verification and performance evaluation of this architecture.

# Appendix

An example code for facilitating OTA join of the xDot is detailed.

```
#include "dot_util.h"

#include "RadioEvent.h"


#if ACTIVE_EXAMPLE == OTA_EXAMPLE

        ////////////////////////////////////////////////////////////////////////////////

        // -------------------- DOT LIBRARY REQUIRED ------------------------------//

        // * Because these example programs can be used for both mDot and xDot     //

        //     devices, the LoRa stack is not included. The libmDot library should //

        //     be imported if building for mDot devices. The libxDot library       //

        //     should be imported if building for xDot devices.                    //

        // * https://developer.mbed.org/teams/MultiTech/code/libmDot-dev-mbed5/    //

        // * https://developer.mbed.org/teams/MultiTech/code/libmDot-mbed5/        //

        // * https://developer.mbed.org/teams/MultiTech/code/libxDot-dev-mbed5/    //

        // * https://developer.mbed.org/teams/MultiTech/code/libxDot-mbed5/        //

        ////////////////////////////////////////////////////////////////////////////////

        //////////////////////////////////////////////////////////////

        // * these options must match the settings on your gateway //

        // * edit their values to match your configuration         //

        // * frequency sub band is only relevant for the 915 bands //

        // * either the network name and passphrase can be used or //

        //     the network ID (8 bytes) and KEY (16 bytes)         //

        //////////////////////////////////////////////////////////////
```

```
//static std::string network_name = "MultiTech";
//static std::string network_passphrase = "MultiTech";

static uint8_t network_id[] = { 0xEC, 0xFA, 0xF4, 0xFE, 0x00, 0x00, 0x00, 0x01 };
static uint8_t network_key[] = { 0x40, 0x0A, 0xE0, 0xBE, 0x6A, 0x07, 0x61, 0x10, 0x6E, 0x0C,
0x6C, 0x24, 0x04, 0xC1, 0x51, 0x41 };
static uint8_t frequency_sub_band = 1;
static lora::NetworkType network_type = lora::PUBLIC_LORAWAN;
static uint8_t join_delay = 5;
static uint8_t ack = 0;
static bool adr = true;

// deepsleep consumes slightly less current than sleep
// in sleep mode, IO state is maintained, RAM is retained, and application will resume after
waking up
// in deepsleep mode, IOs float, RAM is lost, and application will start from beginning after
waking up
// if deep_sleep == true, device will enter deepsleep mode

static bool deep_sleep = false;

mDot* dot = NULL;

lora::ChannelPlan* plan = NULL;
Serial pc(USBTX, USBRX);
```

```
#if defined(TARGET_XDOT_L151CC)
	I2C i2c(I2C_SDA, I2C_SCL);
	ISL29011 lux(i2c);

#else
	AnalogIn lux(XBEE_AD0);
#endif

int main() {
	// Custom event handler for automatically displaying RX data
	RadioEvent events;
	pc.baud(115200);

	#if defined(TARGET_XDOT_L151CC)
		i2c.frequency(400000);
	#endif

	mts::MTSLog::setLogLevel(mts::MTSLog::TRACE_LEVEL);


	#if CHANNEL_PLAN == CP_US915
		plan = new lora::ChannelPlan_US915();

	#elif CHANNEL_PLAN == CP_AU915
		plan = new lora::ChannelPlan_AU915();
```

```
#elif CHANNEL_PLAN == CP_EU868
	plan = new lora::ChannelPlan_EU868();

#elif CHANNEL_PLAN == CP_KR920
	plan = new lora::ChannelPlan_KR920();

#elif CHANNEL_PLAN == CP_AS923
	plan = new lora::ChannelPlan_AS923();

#elif CHANNEL_PLAN == CP_AS923_JAPAN
	plan = new lora::ChannelPlan_AS923_Japan();

#elif CHANNEL_PLAN == CP_IN865
	plan = new lora::ChannelPlan_IN865();

#endif

assert(plan);

dot = mDot::getInstance(plan);

assert(dot);

// attach the custom events handler
dot->setEvents(&events);
```

```
if (!dot->getStandbyFlag()) {
	logInfo("mbed-os library version: %d", MBED_LIBRARY_VERSION);
	//start from a well-known state
	logInfo("defaulting Dot configuration");
	dot->resetConfig();
	dot->resetNetworkSession();
	// make sure library logging is turned on
	dot->setLogLevel(mts::MTSLog::INFO_LEVEL);
	// update configuration if necessary
	if (dot->getJoinMode() != mDot::OTA) {
		logInfo("changing network join mode to OTA");
		if (dot->setJoinMode(mDot::OTA) != mDot::MDOT_OK) {
			logError("failed to set network join mode to OTA");
		}
	}

	// in OTA and AUTO_OTA join modes, the credentials can be passed to the
	library as a name and passphrase or an ID and KEY
	// only one method or the other should be used!
	// network ID = crc64(network name)
	// network KEY = cmac(network passphrase)
	//update_ota_config_name_phrase(network_name, network_passphrase,

	frequency_sub_band, network_type, ack);
	update_ota_config_id_key(network_id, network_key, frequency_sub_band,
	network_type, ack);
```

```
// configure network link checks
// network link checks are a good alternative to requiring the gateway to ACK
every packet and should allow a single gateway to handle more Dots
// check the link every count packets
// declare the Dot disconnected after threshold failed link checks
// for count = 3 and threshold = 5, the Dot will ask for a link check response
every 5 packets and will consider the connection lost if it fails to receive 3
responses in a row

update_network_link_check_config(3, 5);

// enable or disable Adaptive Data Rate
dot->setAdr(adr);

// Configure the join delay
dot->setJoinDelay(join_delay);

// save changes to configuration
logInfo("saving configuration");

if (!dot->saveConfig()) {
        logError("failed to save configuration");
}
```

```
        // display configuration
        display_config();
    }
    else {
        // restore the saved session if the dot woke from deepsleep mode
        // useful to use with deepsleep because session info is otherwise lost when the dot
        enters Deepsleep
        logInfo("restoring network session from NVM");
        dot->restoreNetworkSession();
    }

    while (true) {
        uint16_t light;
        std::vector<uint8_t> tx_data;

        // join network if not joined
        if (!dot->getNetworkJoinStatus()) {
            join_network();
        }

        #if defined(TARGET_XDOT_L151CC)
            // configure the ISL29011 sensor on the xDot-DK for continuous ambient
            light sampling, 16 bit conversion, and maximum range
            lux.setMode(ISL29011::ALS_CONT);
            lux.setResolution(ISL29011::ADC_16BIT);
            lux.setRange(ISL29011::RNG_64000);
```

```
        // get the latest light sample and send it to the gateway
        light = lux.getData();
        tx_data.push_back((light >> 8) & 0xFF);
        tx_data.push_back(light & 0xFF);
        logInfo("light: %lu [0x%04X]", light, light);
        send_data(tx_data);

        // put the LSL29011 ambient light sensor into a low power state
        lux.setMode(ISL29011::PWR_DOWN);

#else
        // get some dummy data and send it to the gateway
        light = lux.read_u16();
        tx_data.push_back((light >> 8) & 0xFF);
        tx_data.push_back(light & 0xFF);
        logInfo("light: %lu [0x%04X]", light, light);
        send_data(tx_data);

#endif

// if going into deepsleep mode, save the session so we don't need to join again
after waking up
// not necessary if going into sleep mode since RAM is retained
if (deep_sleep) {
        logInfo("saving network session to NVM");
```

```
                    dot->saveNetworkSession();
                }

                // ONLY ONE of the three functions below should be uncommented depending
                on the desired wakeup method
                //sleep_wake_rtc_only(deep_sleep);
                //sleep_wake_interrupt_only(deep_sleep);
                sleep_wake_rtc_or_interrupt(deep_sleep);
            }

            return 0;
        }
#endif
```

# Related Publications

- **R. Muppala**, A. Navnit, D. Devendra, E. R. Matera, N. Accettura, and A. M. Hussain, "Feasibility of standalone TDoA-based localization using LoRaWAN", in *Proc. 11th Int. Conf. Localization GNSS (ICL-GNSS)*, Jun. 2021, pp. 1–7.
- **R. Muppala**, A. Navnit, S. Poondla, and A. M. Hussain, "Investigation of indoor LoRaWAN signal propagation for real-world applications," in *Proc. 6th Int. Conf. Converg. Technol. (I2CT)*, 2021, pp. 1–5.

# Bibliography

[1] A. Zanella, N. Bui, A. Castellani, L. Vangelista and M. Zorzi, “Internet of Things for smart cities”, *IEEE Internet Things J.,* vol. 1, no. 1, pp. 22-32, Feb. 2014.

[2] Y. Mehmood, F. Ahmad, I. Yaqoob, A. Adnane, M. Imran and S. Guizani, "Internet-of-Things-based smart cities: Recent advances and challenges", *IEEE Commun. Mag.,* vol. 55, no. 9, pp. 16-24, Sep. 2017.

[3] J. Jin, J. Gubbi, S. Marusic and M. Palaniswami, "An information framework for creating a smart city through Internet of Things", *IEEE Internet Things J.,* vol. 1, no. 2, pp. 112-121, Apr. 2014.

[4] L. Da Xu, W. He and S. Li, “Internet of Things in industries: A survey”, *IEEE Trans. Ind. Informat.,* vol. 10, no. 4, pp. 2233-2243, Nov. 2014.

[5] K. Tange, M. De Donno, X. Fafoutis and N. Dragoni, "A systematic survey of industrial Internet of Things security: Requirements and fog computing opportunities", *IEEE Commun. Surveys Tuts.,* vol. 22, no. 4, pp. 2489-2520, Jul. 2020.

[6] M. Wollschlaeger, T. Sauter and J. Jasperneite, "The future of industrial communication: Automation networks in the era of the Internet of Things and industry 4.0", *IEEE Ind. Electron. Mag.,* vol. 11, pp. 17-27, Mar. 2017.

[7] S. M. Riazul Islam, D. Kwak, M. Humaun Kabir, M. Hossain and K. S. Kwak, “The Internet of Things for health care: A comprehensive survey”, *IEEE Access,* vol. 3, pp. 678-708, Jun. 2015.

[8] A. M. Hussain, F. A. Ghaffar, S. I. Park, J. A. Rogers, A. Shamim and M. M. Hussain, "Metal/polymer based stretchable antenna for constant frequency far-field communication in wearable electronics", *Adv. Funct. Mater.,* vol. 25, no. 42, pp. 6565-6575, 2015.

[9] J. M. Nassar, J. P. Rojas, A. M. Hussain and M. M. Hussain, "From stretchable to reconfigurable inorganic electronics", *Extreme Mech. Lett.*, vol. 9, pp. 245-268, Dec. 2016.

[10] A. M. Hussain and M. M. Hussain, "Deterministic Integration of Out-of-Plane Sensor Arrays for Flexible Electronic Applications", *Small*, vol. 12, no. 37, pp. 5141-5145, 2016.

[11] H. M. Fahad, A. M. Hussain, G. A. Sevilla Torres, S. K. Banerjee, and M. M. Hussain, “Group IV nanotube transistors for next generation ubiquitous computing”, in *Proc. SPIE Micro and Nanotec. Sens. Syst. and Appl. VI*, vol. 9083, 2014.

[12] A. N. Hanna, A. M. Hussain, H. Omran, S. Alsharif, K. N. Salama and M. M. Hussain, "Zinc oxide integrated wavy channel thin-film transistor-based high-performance digital circuits", *IEEE Electron Device Lett.*, vol. 37, no. 2, pp. 193-196, Feb. 2016.

[13] M. M. Hussain *et al.*, “Free form CMOS electronics: Physically flexible and stretchable”, *IEDM Tech. Dig.,* pp. 19.4.1-19.4.4, Dec. 2015.

[14] S. F. Shaikh, M. T. Ghoneim, G. A. Torres Sevilla, J. M. Nassar, A. M. Hussain and M. M. Hussain, “Freeform Compliant CMOS Electronic Systems for Internet of Everything Applications”, *IEEE Trans. Electron Devices,* vol. 64, no. 5, pp. 1894-1905, May 2017.

[15] A. M. Hussain and M. M. Hussain, “CMOS-technology-enabled flexible and stretchable electronics for Internet of everything applications”, *Adv. Mater.*, vol. 28, no. 22, pp. 4219-4249, Jun. 2016.

[16] R. B. Mishra, W. Babatain, N. El-Atab, A. M. Hussain, and M. M. Hussain, “Polymer/paper-based double touch mode capacitive pressure sensing element for wireless control of robotic arm”, *Proc. 15th IEEE Int. Conf. Nano/Micro Engineered Molecular Syst.*, pp. 95-99, Sep. 2020.
[17] R. B. Mishra, S. F. Shaikh, A. M. Hussain and M. M. Hussain, "Metal coated polymer and paper-based cantilever design and analysis for acoustic pressure sensing", *AIP Adv.*, vol. 10, no. 5, May 2020.
[18] G. A. T. Sevilla, M. T. Ghoneim, H. Fahad, J. P. Rojas, A. M. Hussain and M. M. Hussain, "Flexible nanoscale high-performance FinFETs", *ACS Nano*, vol. 8, no. 10, pp. 9850-9856, Sep. 2014.
[19] J. P. Rojas *et al.*, “Transformational electronics are now reconfiguring”, in *Proc. SPIE Micro and Nanotec. Sens. Syst. and Appl. VII*, vol. 9467, 2015.
[20] G. A. T. Sevilla, A. S. Almuslem, A. Gumus, A. M. Hussain, M. E. Cruz and M. M. Hussain, " High performance high-κ /metal gate complementary metal oxide semiconductor circuit element on flexible silicon ", *Appl. Phys. Lett.*, vol. 108, no. 9, pp. 094102, 2016.
[21] N. Ahmed, D. De and I. Hussain, "Internet of Things (IoT) for smart precision agriculture and farming in rural areas", *IEEE Internet Things J.*, vol. 5, no. 6, pp. 4890-4899, Dec. 2018.
[22] C. R. Reddy *et al.*, “Improving spatio-temporal understanding of particulate matter using low-cost IoT sensors”, in *Proc. IEEE 31st Annu. Int. Symp. Pers. Indoor Mobile Radio Commun. (PIMRC).*, pp. 1-7, 2020.
[23] S. Alletto et al., "An indoor location-aware system for an IoT-based smart museum", *IEEE Internet Things J.*, vol. 3, no. 2, pp. 244-253, Apr. 2016.
[24] A. Augustin, J. Yi, T. Clausen and W. M. Townsley, “A study of LoRa: Long range & low power networks for the Internet of Things”, *Sensors*, vol. 16, no. 9, pp. 1466, 2016.
[25] J. Haxhibeqiri, E. D. Poorter, I. Moerman, and J. Hoebeke, “A Survey of LoRaWAN for IoT: From Technology to Application”, *Sensors*, vol. 18, no. 11, p. 3995, 2018.
[26] R. S. Sinha, Y. Wei, and S.-H. Hwang, “A survey on LPWA technology: LoRa and NB-IoT”, *ICT Exp.*, vol. 3, no. 1, pp. 14-21, 2017.
[27] K. Mikhaylov, J. Petäjäjärvi, and T. Haenninen, “Analysis of capacity and scalability of the LoRa low power wide area network technology”, in *Proc. Eur. Wireless 22nd Eur. Wireless Conf.*, May 2016, pp. 1–6.
[28] M. Centenaro, L. Vangelista, A. Zanella, and M. Zorzi, “Long-range communications in unlicensed bands: the rising stars in the IoT and smart city scenarios”, *IEEE Wireless Commun.*, vol. 23, no. 5, pp. 60–67, Oct 2016.
[29] M. Saravanan, A. Das and V. Iyer, "Smart water grid management using LPWAN IoT technology," *Proc. Global Internet Things Summit (GIoTS)*, pp. 1-6, Jun. 2017.
[30] E. Sisinni, P. Bellagente, A. Depari, P. Ferrari, A. Flammini, S. Marella, M. Pasetti, S. Rinaldi, A. Cagiano, "A new LoRaWAN adaptive strategy for smart metering applications," *Proc. IEEE Int. Workshop on Metrology for Industry 4.0 and IoT*, 2020, pp. 690-695.
[31] R. K. Kodali, K. Y. Borra, S. S. G. N. and H. J. Domma," An IoT Based Smart Parking System Using LoRa", *Proc. Int. Conf. Cyber-Enabled Distrib. Comput. Knowl. Discovery (CyberC)*, 2018, pp. 151-1513.
[32] S. A. A'ssri, F. H. K. Zaman and S. Mubdi, "The efficient parking bay allocation and management system using LoRaWAN", *Proc. IEEE 8th Control Syst. Graduate Res. Colloq. (ICSGRC)*, 2017, pp. 127-131.

[33] D. Davcev, K. Mitreski, S. Trajkovic, V. Nikolovski and N. Koteli, "IoT agriculture system based on LoRaWAN", *Proc. 14th IEEE Int. Workshop Factory Commun. Syst. (WFCS)*, pp. 1-4, Jun. 2018.

[34] Z. A. Pandangan and M. C. R. Talampas, "Hybrid LoRaWAN Localization using Ensemble Learning", *Proc. Global Internet Things Summit (GIoTS)*, 2020, pp. 1-6.

[35] C. D. Fernandes, A. Depari, E. Sisinni, P. Ferrari, A. Flammini, S. Rinaldi, M. Pasetti, "Hybrid indoor and outdoor localization for elderly care applications with LoRaWAN", *Proc. IEEE Int. Workshop Med. Meas. Appl. (MeMeA)*, 2020, pp. 1-6.

[36] J. A. del Peral-Rosado, R. Raulefs, J. A. L´opez-Salcedo, and G. Seco Granados, "Survey of cellular mobile radio localization methods: From 1G to 5G", *IEEE Commun. Surveys Tuts.*, vol. 20, no. 2, pp. 1124–1148, 2nd Quart. 2018.

[37] C. Yang and H. Shao, "WiFi-based indoor positioning", *IEEE Commun. Mag.*, vol. 53, no. 3, pp. 150–157, Mar 2015.

[38] I. Guvenc, C. Chong, and F. Watanabe, "NLOS identification and mitigation for UWB localization systems", in *Proc. IEEE Wireless Commun. Netw. Conf. (WCNC)*, Mar 2007, pp. 1571–1576.

[39] B. Hofmann-Wellenhof, H. Lichtenegger, and E. Wasle, *GNSS — Global Navigation Satellite Systems: GPS GLONASS Galileo & More,*. Vienna: Springer, 2008.

[40] H. Liu, H. Darabi, P. Banerjee, and J. Liu, "Survey of wireless indoor positioning techniques and systems", *IEEE Trans. Syst., Man, Cybern. C*, vol. 37, no. 6, pp. 1067–1080, Nov 2007.

[41] Y. Gao, X. Meng, C. M. Hancock, S. Stephenson and Q. Zhang, "UWB/GNSS-based cooperative positioning method for V2X applications", *Proc. 27th Int. Tech. Meet. Satell. Div. U.S. Inst. Navigat.*, pp. 3212-3221, Sep. 2014.

[42] G. Soatti, M. Nicoli, N. Garcia, B. Denis, R. Raulefs and H. Wymeersch, "Implicit cooperative positioning in vehicular networks", *IEEE Trans. Intell. Transp. Syst.*, vol. 19, no. 12, pp. 3964-3980, Dec. 2018.

[43] J. Xiong, J. Cheong, Z. Xiong, A. G. Dempster, M. List, F. Woske, B. Rievers, "Carrier-Phase-Based Multi-Vehicle Cooperative Positioning Using V2V Sensors", *IEEE Trans. Veh. Technol.*, vol. 69, no. 9, pp. 9528-9541, Sep. 2020.

[44] A. H. Ali, M. R. A. Razak, M. Hidayab, S. A. Azman, M. Z. M. Jasmin and M. A. Zainol, "Investigation of indoor WiFi radio signal propagation", *Proc. IEEE Symp. Ind. Electron. Appl. (ISIEA)*, pp. 117-119, 2010.

[45] H. K. Rath, S. Timmadasari, B. Panigrahi and A. Simha, "Realistic indoor path loss modeling for regular WiFi operations in India", *Proc. 23rd Nat. Conf. Commun. (NCC)*, vol. 1, pp. 1-6, Mar. 2017.

[46] M. Hidayab, A. H. Ali and Abas K. B. Azmi, "Wi-Fi signal propagation at 2.4 GHz", *Asia Pacific Microw. Conf.*, pp. 528-531, Dec. 2009.

[47] X. Zhao, Z. Xiao, A. Markham, N. Trigoni and Y. Ren, "Does BTLE measure up against WiFi? A comparison of indoor location performance", Proc. Eur. Wireless Conf., pp. 1-6, May 2014.

[48] Z. Jianyong, L. Haiyong, C. Zili and L. Zhaohui, "RSSI Based Bluetooth Low Energy Indoor Positioning", *Proc. Int. Conf. Indoor Positioning Indoor Navigat.*, pp. 526-533, Oct. 2014.

[49] A. I. Sulyman, A. T. Nassar, M. K. Samimi, G. R. MacCartney, T. S. Rappaport and A. Alsanie, "Radio propagation path loss models for 5G cellular networks in the 28 GHz and 38 GHz millimeter-wave bands", *IEEE Commun. Mag.*, vol. 52, no. 9, pp. 78-86, Sep. 2014.

[50] T. S. Rappaport, "Broadband Millimeter-Wave Propagation Measurements and Models Using Adaptive-Beam Antennas for Outdoor Urban Cellular Communications", *IEEE Trans. Antennas and Propagation,* vol. 61, no. 4, pp. 1850-59, Apr. 2013.
[51] J. N. Murdock, "A 38 GHz Cellular Outage Study for an Urban Campus Environment", *Proc. IEEE Wireless Commun. Net. Conf.,* pp. 3085-90, Apr. 2012.
[52] W. Li, X. Hu and T. Jiang, "Path loss models for IEEE 802.15.4 vehicle-to-infrastructure communications in rural areas", *IEEE Internet Things J.*, vol. 5, no. 5, pp. 3865-3875, Oct. 2018.
[53] J. A. Nazabal, F. Falcone, C. Fern´andez-Valdivielso and I. R. Matías, "Development of a low mobility IEEE 802.15.4 compliant VANET system for urban environments", *Sensors,* vol. 13, no. 6, pp. 7065-7078, 2013.
[54] M. Hata, "Empirical Formula for Propagation Loss in Land Mobile Radio Services", *IEEE Trans. Vehic. Tech.*, vol. VT-29, no. 3, pp. 317-25, Aug. 1980.
[55] V. Erceg et al., "An empirically based path loss model for wireless channels in suburban environments", *IEEE J. Sel. Areas Commun.*, vol. 17, no. 7, pp. 1205-1211, Jul. 1999.
[56] B. C. Fargas and M. N. Petersen, "GPS-free geolocation using LoRa in low-power WANs", in *Proc. Global Internet Things Summit,* Jun 2017, pp. 1–6.
[57] Q. Zhang, P. Wang, and Z. Chen, "An improved particle filter for mobile robot localization based on particle swarm optimization", *Expert Syst. Appl.*, vol. 135, pp. 181–193, 2019.
[58] I. Ullah, Y. Shen, X. Su, C. Esposito, and C. Choi, "A localization based on unscented kalman filter and particle filter localization algorithms", *IEEE Access,* vol. 8, pp. 2233–2246, 2020.
[59] F. Zafari, A. Gkelias, and K. K. Leung, "A survey of indoor localization systems and technologies", *IEEE Commun. Surveys Tuts.,* vol. 21, no. 3, pp. 2568–2599, 3rd Quart. 2019.
[60] E. Yurtsever, J. Lambert, A. Carballo, and K. Takeda, "A survey of autonomous driving: Common practices and emerging technologies", *IEEE Access,* vol. 8, pp. 58 443–58 469, 2020.
[61] A. T. C. LoRa, *RP002-1.0.1 LoRaWAN® Regional Parameters,* Feb. 2020, rP002-1.0.1.
[62] D. Torrieri, *Principles of Spread-Spectrum Communication Systems,* New York: Springer-Verlag, 2004.
[63] Semtech, *LoRa Modulation Basics AN1200.22,* Rev. 2, May. 2013.
[64] LoRa Alliance, *LoRaWAN™ - What is it?: A technical overview of LoRa® and LoRaWAN*, Technical Marketing Workgroup 1.0, Nov. 2015.
[65] L. Gregora, L. Vojtech and M. Neruda, "Indoor signal propagation of LoRa technology", *Proc. Int. Conf. Mechatronics Mechatronika (ME)*, pp. 1-4, 2016.
[66] J. Petäjäjärvi, K. Mikhaylov, A. Roivainen, T. Hanninen and M. Pettissalo, "On the coverage of LPWANs: Range evaluation and channel attenuation model for LoRa technology", *Proc. ITST,* pp. 55-59, Dec. 2015.
[67] W. Xu, J. Y. Kim, W. Huang, S. S. Kanhere, S. K. Jha and W. Hu, "Measurement characterization and modeling of LoRa technology in multifloor buildings", *IEEE Internet Things J.,* vol. 7, no. 1, pp. 298-310, Jan. 2020.
[68] S. Hosseinzadeh, H. Larijani, K. Curtis, A. Wixted and A. Amini, "Empirical propagation performance evaluation of LoRa for indoor environment", *Proc. IEEE 15th Int. Conf. Ind. Informat. (INDIN),* pp. 26-31, 2017.

[69] R. Muppala, A. Navnit, S. Poondla, and A. M. Hussain, “Investigation of indoor LoRaWAN signal propagation for real-world applications,” in *Proc. 6th Int. Conf. Converg. Technol. (I2CT)*, 2021, pp. 1–5.

[70] Y. Zou and Q. Wan, “Asynchronous time-of-arrival-based source localization with sensor position uncertainties”, *IEEE Commun. Lett.*, vol. 20, no. 9, pp. 1860–1863, Sep. 2016.

[71] N. Patwari, J. N. Ash, S. Kyperountas, A. O. Hero, R. L. Moses, and N. S. Correal, “Locating the nodes: cooperative localization in wireless sensor networks”, *IEEE Signal Process. Mag.*, vol. 22, no. 4, pp. 54–69, July 2005.

[72] Y. Wang, X. Ma, and G. Leus, “Robust time-based localization for asynchronous networks”, *IEEE Trans. Signal Process.*, vol. 59, no. 9, pp. 4397–4410, Sep. 2011.

[73] F. Viani, P. Rocca, M. Benedetti, G. Oliveri, and A. Massa, “Electromagnetic passive localization and tracking of moving targets in a WSN-infrastructured environment”, *Inverse Probl.*, vol. 26, pp. 1–15, Mar 2010.

[74] W. Bakkali, M. Kieffer, M. Lalam, and T. Lestable, “Kalman filter based localization for Internet of Things LoRaWAN™ end points”, in *Proc. IEEE 28th Annu. Int. Symp. Pers. Indoor Mobile Radio Commun. (PIMRC).*, Oct 2017, pp. 1–6.

[75] M. Aernouts, R. Berkvens, K. Van Vlaenderen, and M. Weyn, “Sigfox and LoRaWAN Datasets for Fingerprint Localization in Large Urban and Rural Areas”, *Data*, vol. 3, no. 2, p. 13, Apr 2018.

[76] S. Sadowski and P. Spachos, “RSSI-based indoor localization with the Internet of Things”, *IEEE Access*, vol. 6, pp. 30 149–30 161, 2018.

[77] H. Kwasme and S. Ekin, “RSSI-based localization using LoRAWAN technology”, *IEEE Access*, vol. 7, pp. 99856-99866, 2019.

[78] D. Plets, N. Podevijn, J. Trogh, L. Martens, and W. Joseph, “Experimental Performance Evaluation of Outdoor TDoA and RSS Positioning in a Public LoRa Network”, in *Proc IEEE Int. Conf. Indoor Position. Indoor Navig.*, Sep. 2018, pp. 1–8.

[79] A. A. Ghany, B. Uguen, and D. Lemur, “A pre-processing algorithm utilizing a paired CRLB for TDoA based IoT positioning,” in *Proc. IEEE Veh. Technol. Conf.*, May 2020, pp. 1–5.

[80] M. Aernouts, N. BniLam, N. Podevijn, D. Plets, W. Joseph, R. Berkvens, and M. Weyn, “Combining TDoA and AoA with a particle filter in an outdoor LoRaWAN network,” in *Proc. IEEE/ION Position Location Navigat. Symp.*, Apr 2020, pp. 1060–1069.

[81] Z. Sahinoglu, S. Gezici, and I. G¨uvenc, Ultra-wideband Positioning Systems: Theoretical Limits, Ranging Algorithms, and Protocols. Cambridge University Press, 2008, ch. 4, p. 63–100.

[82] I. D. Coope, "Reliable computation of the points of intersection of n spheres in $R^n$", *The ANZIAM (Australian and New Zealand Industrial and Applied Mathematics) Journal*, vol. 42, no. E, pp. C461-C477, 2000.

[83] R. Muppala, A. Navnit, D. Devendra, E. R. Matera, N. Accettura, and A. M. Hussain, “Feasibility of standalone TDoA-based localization using LoRaWAN”, in *Proc. 11th Int. Conf. Localization GNSS (ICL-GNSS)*, Jun. 2021, pp. 1–7.

[84] H. H. Fan and C. Yan, “Asynchronous differential TDOA for sensor self localization,” in *Proc. IEEE Int. Conf. Acoust. Speech Signal Process.*, vol. 2, 2007, pp. 1109–1112.

[85] C. G. Ramirez, A. Dyussenova, A. Sergeyev, and B. Iannucci, “Long-ShoT: Long-range synchronization of time,” in *Proc. 18th Int. Conf. Inf. Process. Sensor Netw.*, 2019, pp. 289–300.

[86] A. Vashistha and C. L. Law, "E-DTDOA based localization for wireless sensor networks with clock drift compensation," *IEEE Sensors J.*, vol. 20, no. 5, pp. 2648–2658, 2020.
[87] R. Achenbach, M. Feuerstack-Raible, F. Hiller, M. Keller, K. Meier, H. Rudolph, and R. Saur-Brosch, "A digitally temperature-compensated crystal oscillator," *IEEE J. Solid-State Circuits*, vol. 35, no. 10, pp. 1502– 1506, 2000.
[88] T. Haapala, A. Liscidini, and K. A. I. Halonen, "Temperature compensation of crystal references in NB-IoT modems," *IEEE Trans. Circuits Syst. I*, vol. 67, no. 7, pp. 2467–2480, 2020.
[89] Semtech, *LoRa Modem Design Guide: SX1272/3/6/7/8*, Jul. 2013.